\documentclass[prx,aps,amssymb,shownopacs,twocolumn,10pt,floatfix]{revtex4-1}
\usepackage{amsmath}
\usepackage{amssymb}
\usepackage{amsthm}
\usepackage{amsfonts}
\usepackage{listings}
\usepackage{enumerate}
\usepackage{latexsym}
\usepackage{hyperref}
\usepackage{braket}

\usepackage{psfrag}
\usepackage{color}
\usepackage{bm}
\usepackage[pdftex]{graphicx}
\usepackage{subfigure}

\def\ie{i.e.,\ }
\def\eg{e.g.\ }

\def\matrix22#1#2#3#4{\left(\begin{array}{cc}#1&#2\\#3&#4\end{array}\right)}

\usepackage{ulem}

\newcommand{\mbr}{{\mathbf{r}}}
\newcommand{\jain}{\Psi^{[-2,-2]}_{\rm CF}}
\newcommand{\wf}[2]{\Psi^{[#1,#2]}_{\rm CF}}

\newcommand{\up}{{\uparrow}}
\newcommand{\dn}{{\downarrow}}
\newcommand{\norder}[1]{ {\mkern1mu\colon\mkern-4mu{#1}\colon\mkern-3mu} }

\begin{document}

\title{Emergent particle-hole symmetry in spinful bosonic quantum Hall systems}

\author{S. D. Geraedts$^1$}
\author{C. Repellin$^2$}
\author{Chong Wang$^3$ }
\author{Roger S. K. Mong$^4$}
\author{T. Senthil$^5$}
\author{N. Regnault$^6$}
\affiliation{$^1$ Department of Electrical Engineering, Princeton University, Princeton NJ 08544, USA\\
$^2$ Max-Planck-Institut f\"ur Physik komplexer Systeme, 01187 Dresden, Germany\\
$^3$ Department of Physics, Harvard University, Cambridge, Massachusetts, 02138, USA\\
$^4$ Department of Physics and Astronomy, University of Pittsburgh, Pittsburgh, PA 15260, United States\\
$^5$ Department of Physics, Massachusetts Institute of Technology, Cambridge, MA 02139, USA\\
$^6$ Laboratoire Pierre Aigrain, D\'epartement de physique de l'ENS, \'Ecole normale sup\'erieure, PSL Research University, Universit\'e Paris Diderot, Sorbonne Paris Cit\'e, Sorbonne Universit\'es, UPMC Univ.~Paris 06, CNRS, 75005 Paris, France}

\begin{abstract}
When a fermionic quantum Hall system is projected into the lowest Landau level, there is an exact particle-hole symmetry between filling fractions $\nu$ and $1-\nu$. 
We investigate whether a similar symmetry can emerge in bosonic quantum Hall states, where it would connect states at filling fractions $\nu$ and $2-\nu$.
We begin by showing that the particle-hole conjugate to a composite fermion `Jain state' is another Jain state, obtained by reverse flux attachment. We show how information such as the shift and the edge theory can be obtained for states which are particle-hole conjugates.
Using the techniques of exact diagonalization and infinite density matrix renormalization group, we study a system of two-component (i.e., spinful) bosons, interacting via a $\delta$-function potential. We first obtain real-space entanglement spectra for the bosonic integer quantum Hall effect at $\nu=2$, which plays the role of a filled Landau level for the bosonic system. 
We then show that at $\nu=4/3$ the system is described by a Jain state which is the particle-hole conjugate of the Halperin (221) state at $\nu=2/3$. We show a similar relationship between non-singlet states at $\nu=1/2$ and $\nu=3/2$.
We also study the case of $\nu=1$, providing unambiguous evidence that the ground state is a composite Fermi liquid.
Taken together our results demonstrate that there is indeed an emergent particle-hole symmetry in bosonic quantum Hall systems.
\end{abstract}

\date{\today}
\maketitle

\section{Introduction}

Pariticle-hole symmetry arises in many electronic systems, such as graphene and Weyl semimetals, where electrons and holes behave alike.
As electron-electron interactions (i.e.,~Coulomb repulsion) respect particle-hole symmetry, only the band structure and chemical potential must be tuned to achieve symmetry.
Recently, there has been a resurgence of interest in the role of particle-hole symmetry in quantum Hall systems, for which the symmetry exists naturally without any fine-tunning.
The magnetic field quenches the kinetic energy and particle-hole becomes an exact microscopic symmetry within a single Landau level.
The role of particle-hole symmetry at half-filling, in particular the possible Dirac nature of the composite fermions (CF)~\cite{Son-PhysRevX.5.031027,Wang-PhysRevX.5.041031,dualdrmaxav,kachru15,Wang-PhysRevB.93.085110,Geraedts197,Murthy-PhysRevB.93.085405,dualdrMAM}, has forced a reevaluation of the established theory of the composite Fermi liquid phase.
While the Dirac CFs would immediately lead to a particle-hole symmetric CF Fermi liquid at $\nu=1/2$, it seems that such a feature might also hold true for non-relativistic CFs~\cite{Wang-2017arXiv170100007W}. 
Thinking about particle-hole symmetry has also revealed deep connections between the quantum Hall effect and other topological phases (including topological insulators and gapless spin liquids)~\cite{tsymmu1,Wang-PhysRevX.5.041031,dualdrmaxav,Wang-PhysRevB.93.085110,Metlitski-2015arXiv151005663M}, as well as with `duality webs' often discussed in the high-energy physics~\cite{Seiberg2016,KarchTong,Murugan,kachrubosonization}.

It is natural to consider if such ideas extends to the bosonic case, where the microscopic constituents no longer obey Pauli exclusion principle.
There has been much experimental progress in realizing the quantum Hall effect in cold atoms~\cite{Miyake-PhysRevLett.111.185302,Aidelsburger-PhysRevLett.111.185301,jotzu2014experimental,aidelsburger2015measuring} and optical cavities~\cite{mittal2016measurement,schine2016synthetic}.
The fermionic implementation of particle-hole symmetry--the interchange of filled and empty orbitals--cannot be applied to bosonic systems in any obvious way.
Unlike fermions, for which interchanging the creation and annihilation operators preserves their anticommutation relations $\{f, f^\dag\} = 1$, the fundamental commutation relation for bosons $[b,b^\dag] = 1$ is violated upon this exchange.
Nevertheless some of the recently developed ideas for the fermionic particle-hole symmetric CF Fermi liquid state can be extended to the bosonic case at $\nu=1$. 
Two of us~\cite{Wang-PhysRevB.94.245107} have recently suggested that this symmetry could be emergent in the bosonic FQHE, at low energy and long wavelength, close to filling fraction $\nu=1$ (see also Ref.~\onlinecite{Murthy-PhysRevB.93.085405,mross2pi}).

To see how this could be possible, consider that for fermions the particle-hole symmetry in a single Landau level can be thought as a condensation of the $\nu=1$ integer quantum Hall hole excitations into a fractional state (such as the Laughlin state). To extend this symmetry to bosonic systems, it was suggested to use the same construction, substituting the $\nu=1$ integer quantum Hall state with the bosonic integer quantum Hall effect (bIQHE) at $\nu=2$. The bIQHE is the prototype of a bosonic symmetry protected topological phase in two dimensions (see Ref.~\onlinecite{Senthil-annurev-conmatphys-031214-014740} for a short review). The physical properties of this state have been studied in Refs.~\onlinecite{Lu-PhysRevB.86.125119,Senthil-PhysRevLett.110.046801,Liu-PhysRevLett.110.067205,Geraedts2013288}. Following the proposal of Ref.~\onlinecite{Senthil-PhysRevLett.110.046801}, numerical evidence of this phase has been recently found in various microscopic models~\cite{Furukawa-PhysRevLett.111.090401,Wu-PhysRevB.87.245123,Regnault-PhysRevB.88.161106,Moller:2009p184,Sterdyniak-PhysRevLett.115.116802,He-PhysRevLett.115.116803,Zeng-PhysRevB.93.195121}.

In this article, we provide convincing evidence of an emergent particle-hole symmetry in a spinful bosonic quantum Hall system. For that purpose, we use a combination of exact diagonalization on the sphere and torus geometries and iDMRG calculations on the infinite cylinder geometry~\cite{Zaletel-PhysRevLett.110.236801}.

We first review in Sec.~\ref{Sec:ModelAndCFStates} the microscopic spinful bosonic quantum Hall system that we consider and the construction of the bosonic Jain singlet composite fermion states that are relevant for this setup. We use the entanglement spectrum~\cite{Li-PhysRevLett.101.010504} (ES) as a method to characterize the various topological orders. Through the bulk-edge correspondence, the ES allows to extract the edge excitations from the bulk wavefunction. For this reason, we discuss the edge mode structure of the bosonic Jain singlet CF states. In Sec.~\ref{Sec:EmergentPHSymmetry} we then explicitly show how the procedure of Ref.~\onlinecite{Wang-PhysRevB.94.245107} can be used to construct the particle-hole conjugate of a bosonic state. Using this method, we are able to find a relationship between the shift of a quantum Hall state on the sphere and its particle-hole conjugate, a relationship which we will later use to identify particle-hole conjugate states.

\begin{figure}
\centerline{\includegraphics[width=0.92\linewidth]{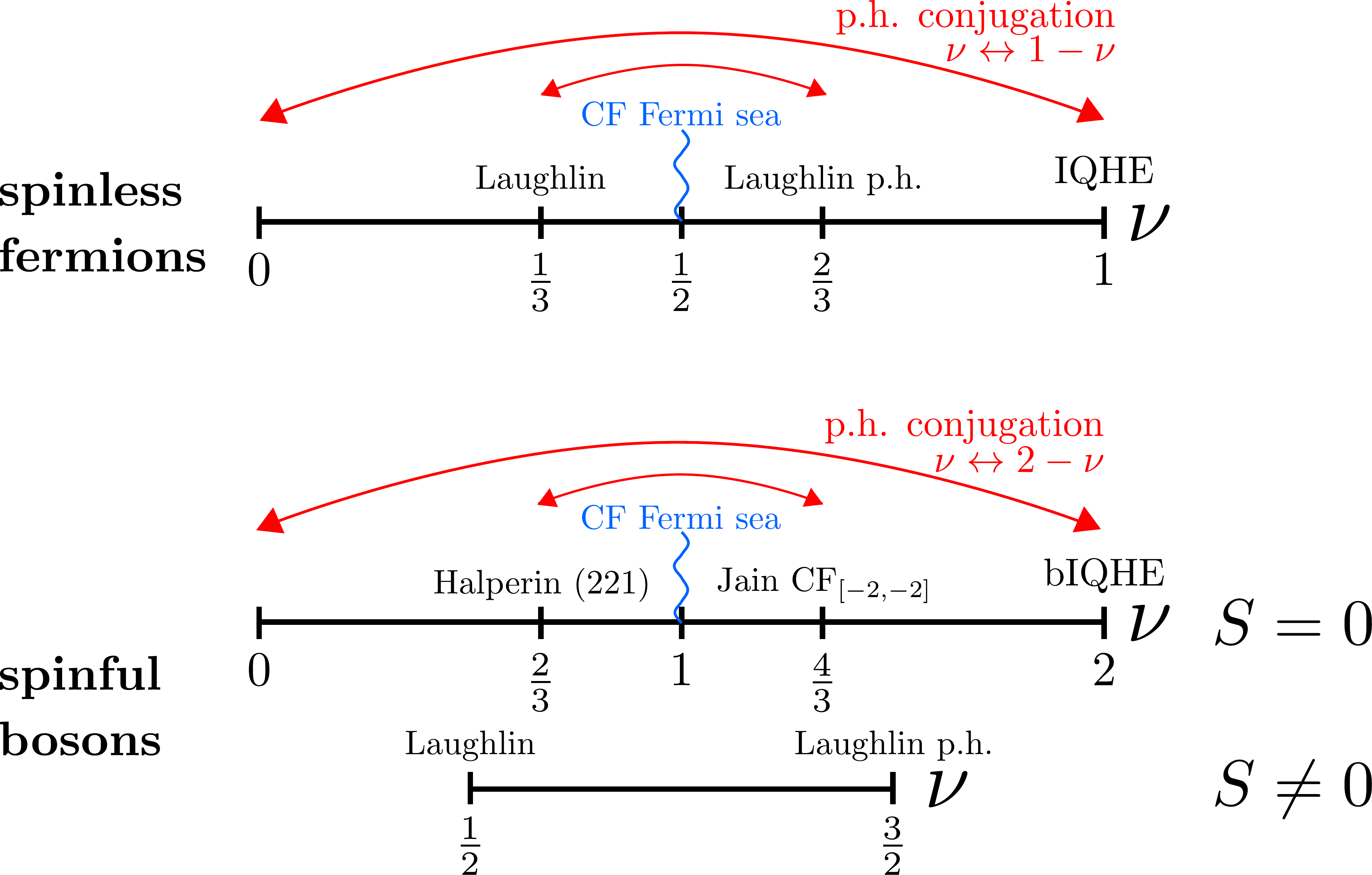}}
\caption{A schematic phase diagram for both spinless fermions and spinful bosons, both in the lowest Landau level and with a typical short range interaction. For spinless fermions, the exact particle hole symmetry relates the vacuum ($\nu=0$) to the integer quantum Hall effect ($\nu=1$) and the Laughlin $\nu=1/3$ state to its particle hole conjugate at $\nu=2/3$. At half filling, a composite fermion Fermi sea arises. The emergent particle hole symmetry in a spinful bosonic quantum Hall system leads to a similar phase diagram where the fermionic integer quantum Hall effect at $\nu=1$ is replaced by the bosonic integer quantum Hall effect at $\nu=2$. The Halperin $(221)$ state is the analogue of the Laughlin state and its particle hole conjugate is described by the bosonic spinful CF state $\Psi^{[-2, -2]}_{\rm CF}$. The p.h. invariant filling factor under this emergent symmetry is now at $\nu=1$ instead of $\nu=1/2$ for spinless fermions, but it develops a similar phase described by a spinful composite fermion Fermi sea. While we mainly focus on spin singlet states ($S=0$), the particle-hole conjugation can be extended to state with a finite spin ($S \neq 0$) such as the $\nu=1/2$ Laughlin state.}
\label{fig:PhaseDiagram}
\end{figure}

A number of filling fractions take on special meaning when particle-hole symmetry is present. In the top part of Fig.~\ref{fig:PhaseDiagram} we give an example of this for the fermionic case which has the following states: (a) a state at $\nu=1$ with respect to which the particle-hole symmetry is performed, (b) a pair of states at $\nu$ and $1-\nu$ (e.g.~$1/3$ and $2/3$) which are related by particle hole symmetry, and (c) a state at $\nu=1/2$, which may be its own particle-hole conjugate~\cite{Son-PhysRevX.5.031027,Wang-PhysRevB.93.085110,Geraedts197,Barkeshli-PhysRevB.92.165125,Wang-2017arXiv170100007W}.

In the bottom part of Fig.~\ref{fig:PhaseDiagram} we show the equivalent picture for bosons. At $\nu=2$, we expect to find the bIQHE phase. In Sec.~\ref{Sec:Nu2} we present a iDMRG study of this phase in our continuum model. The key test of the emergent particle-hole symmetry is a correspondence between states at filling fraction $\nu$ and $2-\nu$. In Sec.~\ref{Sec:Nu43} we demonstrate this correspondence for the Halperin $(221)$ state at $\nu=2/3$ and the $\jain$ state at $\nu=4/3$.
Sec.~\ref{Sec:Nu32} extends this correspondence to non-spin singlet states with the example of $\nu=1/2$ and $\nu=3/2$. 
Finally at $\nu=1$ we find unambiguous evidence that spinful boson physics at this filling factor is indeed a composite Fermi liquid (CFL) \cite{Read-PhysRevB.58.16262}.

%%%%%%%%%%%%%%%%%%%%%%%%%%%%%%%%%%%%%%%%%%%%%%%%%%%%%%%%%%%%%%%%%%
\section{Microscopic model and bosonic spinful CF states}\label{Sec:ModelAndCFStates}

In this work we consider two species of bosons, projected to the lowest Landau level and interacting with the following potential:
\begin{align}
	H = \frac12 \int d^2\mbr \, d^2\mbr' \!\sum_{\sigma,\sigma'=\up,\dn}\! \rho_\sigma(\mbr) \rho_{\sigma'}(\mbr') V_{\sigma\sigma'}(\mbr-\mbr'),
\end{align}
where $\rho_\sigma(\mbr)$ is the boson density. The index $\sigma=\{\up,\dn\}$ stands for the two different species which can be thought as any 2-component internal degree of freedom, layer index,etc. For sake of simplicity, we will use the name spin for this degree of freedom. We restrict ourselves to potentials $V_{\sigma\sigma'}(r)$ independent of the spin indices \ie exhibiting an $SU(2)$ symmetry and drop the spin index for the interaction. Most of our work focuses on the repulsive hardcore interaction $V(\mbr)=\delta^{(2)}(\mbr)$. It is both the simplest and more realistic interaction since the $s$-wave scattering correctly describes cold gases of alkali atoms such as $^{87}\mathrm{Rb}$. 
Also, previous works have found a bIQHE~\cite{Furukawa-PhysRevLett.111.090401,Wu-PhysRevB.87.245123,Regnault-PhysRevB.88.161106} at $\nu=2$ for this interaction, and some evidence which points to a CFL~\cite{WuCFL} at $\nu=1$. Therefore it makes sense to use it as a starting point for our numerical study.
For more generic interactions, we express our potentials in terms of Haldane pseudo-potentials~\cite{Haldane-PhysRevLett.51.605} $V_i$. The hardcore interaction corresponds to the case where only the $V_0$ pseudo-potential is non-zero. The stability of a phase can be probed by adding some longer range interaction such as the $V_1$ pseudo-potential.
\footnote{In principle a phase diagram like that of Fig.~\ref{fig:PhaseDiagram}(b) can also be realized for a single-component (spin-polarized) system. We have attempted to study such a system a variety of fractions, but have been unable to find numerical evidence for the bIQHE or CFL phases, or evidence of an emergent particle-hole symmetry.}

For a spinful bosonic FQH, several model wave functions are relevant to explain possible emerging topological phases. The Halperin state~\cite{Halperin83} is the generalization of the Laughlin state to multi-component particles. Moreover, the hardcore interaction is the model interaction for the Halperin $(221)$ state, a spin singlet state which occurs at filling factor $\nu=2/3$. On the sphere geometry, any model state appears at a specific filling factor $\nu$ and also a particular shift $\delta$ that relate the number of particles $N$ and the number of flux quanta $N_{\Phi}$ through 
\begin{eqnarray}
N_{\Phi}=\nu^{-1} N - \delta.
	\label{eq:ShiftFillingSphere}
\end{eqnarray}
For the Halperin $(221)$ state, the shift is $\delta=2$. A series of non-Abelian spin singlet~\cite{Ardonne-PhysRevLett.82.5096} (NASS) states can be built at filling $\nu=\frac{2 k}{3}$ (and $\delta=2$) by symmetrizing $k$ copies of the Halperin $(221)$. They are the natural extension of the Read-Rezayi series~\cite{Read-1999PhRvB..59.8084R} to the spinful case and are described by the $SU(3)_k$ algebra.

Jain's composite fermions~\cite{Jain-PhysRevB.40.8079} construction can be generalized to obtain a series of spin singlet composite fermion wavefunctions either for fermions~\cite{Wu-PhysRevLett.71.153,Park-PhysRevLett.80.4237,Park-PhysRevLett.83.5543,Davenport-PhysRevB.85.245303} or bosons~\cite{Wu-PhysRevB.87.245123}. These states can be written in a similar fashion to the spinless case. The following wavefunctions correspond respectively to direct and reverse flux attachment:
\begin{align}
	\Psi^{[n, n]}_{\rm CF}\big(\{z\}\big) & =  {\mathcal P}_{LLL} \Big[ \Phi_n\big(\{z^{\uparrow}\}\big) \, \Phi_n\big(\{z^{\downarrow}\}\big) \, J\big(\{z\}\big) \Big]
	\label{eq:spinfulCF}
\end{align}
where $\{z^{\uparrow}\}$ (resp. $\{z^{\downarrow}\}$) are the complex coordinates of the particles with a spin up (resp. down). $J\left(\{z\} \right)$ is the Jastrow factor for all particles. If $n$ is positive, $\Phi_n$ is the Slater determinant representing $n$ filled Landau (or Lambda) levels for the CF. If $n$ is negative $\Phi_{n}$ represents $|n|$ filled Landau (or Lambda) levels with an opposite magnetic field. The Jain state $\Psi^{[n, n]}_{\rm CF}$ appears at $\nu = 2n/(2n+1)$ and has a shift $\delta=1+n$. Note that the Halperin $(221)$ state is identical to $\Psi^{[1, 1]}$. $\Psi^{[-1, -1]}_{\rm CF}$ is a plausible candidate wavefunction~\cite{Senthil-PhysRevLett.110.046801} for the bIQHE at $\nu = 2$ with a relatively good overlap for small system sizes on the sphere geometry~\cite{Wu-PhysRevB.87.245123}. We can also consider a different number of Lambda levels for the spin up and the spin down, i.e., $\Psi^{[\pm n_\uparrow, \pm n_\downarrow]}_{\rm CF}$  at filling $\nu=(n_\uparrow+n_\downarrow)/(n_\uparrow + n_\downarrow \pm 1)$ and shift $\delta= 1 \pm \frac{n_\uparrow^2 + n_\downarrow^2}{n_\uparrow + n_\downarrow}$. These states are still $SU(2)$ eigenstates but are not spin singlets if $n_\uparrow \ne n_\downarrow$. 

The edge mode theory of these singlet composite fermion states is described by their ($2|n|\times 2|n|$ dimensional) $K$ matrices
\begin{equation}
\label{Kn}
K^{[n,n]}=\begin{pmatrix}
          1 & 1 & \cdots & 1 \\
          1 & 1 & \cdots & 1 \\
          \vdots & \vdots & \ddots & \vdots \\
          1 & 1 & \cdots & 1 \end{pmatrix} + \operatorname{sgn}(n)I_{2|n|\times 2|n|},
\end{equation}
where the first term is a $2|n|\times 2|n|$ matrix of which every element is $1$, and $I_{2|n|\times 2|n|}$ is a $2|n|\times 2|n|$ identity matrix. (See Appendix~\ref{KmatrixAppendix} for a derivation). The charges carried by the edge modes can be described by the charge vector, specified separately for up and down spins:
\begin{equation}
\tau_{\up}=\left(\begin{array}{c}
                         1 \\
                        0 \\
                         1 \\
                         0 \\
                         \vdots \\
                         1 \\
                         0 \end{array}\right),\quad
	\tau_{\downarrow}=\left(\begin{array}{c}
                         0 \\
                        1 \\
                         0 \\
                         1 \\
                         \vdots \\
                         0 \\
                         1 \end{array}\right),
\end{equation}
where each vector has $n$ non-vanishing entries. Such a $K$ matrix allows us to deduce the number of edge modes and their chirality. For $n>0$, we find $2n$ propagating (complex) edge modes. For $n<0$, we find a single propagating edge mode (the charge mode) and $2|n|-1$ counter-propagating edge modes. This edge mode structure can be conveniently extracted from the bulk wavefunction through the entanglement spectrum~\cite{Li-PhysRevLett.101.010504} and more particularly from the real space entanglement spectrum~\cite{Dubail-PhysRevB.85.115321,Sterdyniak-PhysRevB.85.125308,Rodriguez-PhysRevLett.108.256806} when counter-propagarting edge modes are present. We show a few examples for concreteness: for $n=1$ we have
\begin{equation}
\label{eq:KMatrixNu23}
K^{[1,1]}=\left(\begin{array}{cc}
         2 & 1 \\
         1 & 2 \end{array}\right),
\end{equation}
which is nothing but the Halperin $(221)$ state at $\nu=2/3$. For $n=-2$ we have
 \begin{equation}
 \label{eq:KMatrixNu43}
K^{[-2,-2]}=\left(\begin{array}{cccc}
          0 & 1 & 1 & 1 \\
          1 & 0 & 1 & 1 \\
          1 & 1 & 0 & 1 \\
          1 & 1 & 1 & 0 \end{array}\right),
\end{equation}
which describes the Jain state at $\nu=4/3$. For $n=-1$ we have
\begin{equation}\label{eq:KMatrixNu2}
K^{[-1,-1]}=\left(\begin{array}{cc}
         0 & 1 \\
         1 & 0 \end{array}\right),
\end{equation}
which is simply the bosonic integer quantum hall state, as expected.

\section{Emergent particle hole symmetry}\label{Sec:EmergentPHSymmetry}

In a spinless fermionic system, particle-hole symmetry can be trivially implemented by  swapping the occupation of filled and empty orbitals. 
When focusing on the filling factors $\nu < 1$, p.h.~symmetry is a robust description of the fractional quantum Hall (FQH) physics, though it can be broken by perturbations such as Landau level mixing~\cite{Bishara-PhysRevB.80.121302, Simon-PhysRevB.87.155426,Peterson-PhysRevLett.113.086401,Sodemann-PhysRevB.87.245425}. This latest ingredient is important to understand the emergence of the non-Abelian states, \eg the Pfaffian~\cite{Moore1991362} and anti-Pfaffian~\cite{Levin-aPf,Lee-aPf} states at filling factor $\nu=5/2$~\cite{Willett:FQH5/2:1987, Eisenstein2002-5/2,Pan1999-5/2-12/5,Xia_12_5}, or the absence of a Hall conductance quantization~\cite{Mong-2015arXiv150502843M, Pakrouski-PhysRevB.94.075108} at $\nu=13/5$ despite the clear experimental signatures of an incompressible state at $\nu=12/5$~\cite{Xia_12_5, Kumar:12fifths, ChoiPhysRevB.77.081301, PanPhysRevB.77.075307}.

The approach of swapping filled and empty orbitals clearly breaks down when we consider a bosonic system since bosons can condense in a single orbital. Recently, Ref.~\onlinecite{Wang-PhysRevB.94.245107} has proposed a different route to define a particle-hole conjugation for bosonic states. Indeed the particle-hole conjugate of a fermionic state can be thought as a condensate of the hole excitations of a filled Landau level into a fractional state. In first-quantized notation, this transformation can be written as
\begin{align}\begin{split}
	\tilde{\Psi}(w_1,\dots,w_M) &= 
		\int dz_1 \cdots dz_N \, \Psi(z_1,\dots,z_N)^\ast
		\\ &\quad \times \Psi_{\rm IQH}(w_1,\dots,w_M;z_1,\dots,z_N) ,
	\label{eq:GenericPsi0ParticleHole}
\end{split}\end{align}
where $\Psi_{\rm IQH}$ is the filled lowest Landau level with $M$ particles at coordinates $w_1,\dots,w_M$ and $N$ (quasi)holes at positions $z_1,\dots,z_N$. The wavefunction $\Psi$ is the state we want take the particle-hole conjugate of. Substituting the filled Landau level $\Psi_{\rm IQH}$ with a generic fractional quantum Hall state will instead lead to towers of hierarchy states, such as the Haldane-Halperin hierarchy~\cite{Haldane-PhysRevLett.51.605,Halperin83,Halperin-PhysRevLett.52.1583} or the hierarchy~\cite{Bonderson-PhysRevB.78.125323} on top of the Moore-Read state~\cite{Moore1991362}.  This construction is directly connected to the CF description of quantum Hall~\cite{Hansson-PhysRevLett.102.166805,Hansson-PhysRevB.80.165330,Suorsa-PhysRevB.83.235130,Bonderson-PhysRevLett.108.066806}.

The expression Eq.~\eqref{eq:GenericPsi0ParticleHole} is valid both for fermions and bosons as long as $\Psi_{\rm IQH}$ and $\Psi$ have the same statistics. Thus a suitable choice for $\Psi_{\rm IQH}$ could be used as the seed of a generalized particle-hole conjugation for bosons. As suggested in Ref.~\onlinecite{Wang-PhysRevB.94.245107}, a natural candidate for the  $\Psi_{\rm IQH}$ is the bosonic integer quantum Hall (bIQH) wavefunction $\Psi_{\rm bIQH}$ at $\nu=2$, the analogue of a fermionic filled Landau level. This leads to a new particle-hole conjugate state $\tilde{\Psi}_B$, at filling fraction $\tilde{\nu}=2-\nu$, for any state $\Psi_B$ at filling fraction $\nu$. 

A natural question is: in what sense can one think of the transformation Eq.~\eqref{eq:GenericPsi0ParticleHole} as a symmetry? For a generic microscopic Hamiltonian (without fine tuning), one clearly does not expect it to be an exact symmetry -- it certainly will not transform the exact ground state at filling $\nu$ to the exact ground state at $2-\nu$. However, it was argued in Refs.~\onlinecite{Murthy-PhysRevB.93.085405,Wang-PhysRevB.94.245107,Wang-2017arXiv170100007W}, that if the microscopic interaction stabilizes a composite fermi liquid (CFL) phase at $\nu=1$, this particle-hole symmetry will emerge as a low-energy, long-wavelength property close to $\nu=1$. In particular, the ground state at filling fraction $\nu$ close to $1$ will be related to the ground state at $2-\nu$ through the particle-hole transform. Therefore the appearance of a CFL at $\nu=1$ implies particle-hole symmetric behavior, at least near $\nu=1$. Conversely, if particle-hole symmetric behavior is observed away from $\nu=1$, then a CFL is likely (but not necessarily) stabilized at $\nu=1$.

Note that we have omitted the internal degree of freedom index in Eq.~\eqref{eq:GenericPsi0ParticleHole}. They should be taken into account for spinful bosons. In particular, $\Psi_{\rm bIQH}$ as built from the CF construction (\ie $\Psi^{[-1, -1]}_{\rm CF}$) is a spin singlet state. So $\Psi_B$ and $\tilde{\Psi}_B$ have the same total spin. We can deduce the filling factor and the shift on the sphere from Eq.~\eqref{eq:GenericPsi0ParticleHole} by noting that the number of flux quanta on both sides are equal. Denoting $\nu$ and $\delta$ (resp.~$\tilde{\nu}$ and $\tilde{\delta}$) the filling factor and the shift of $\Psi_B$ (resp.~$\tilde{\Psi}_B$), we obtain the following relations
\begin{eqnarray}
\tilde{\nu}= 2 - \nu&\;\;\;{\rm and}\;\;\;&\tilde{\nu}\tilde{\delta}=-\nu\delta\label{eq:FillingShiftBosonParticleHoleBIQHE}.
\end{eqnarray}
We immediately see that these relations are also satisfied when considering the spinful CF states $\Psi_B=\Psi^{[n, n]}_{\rm CF}$ and $\tilde{\Psi}_B=\Psi^{[-(n+1), -(n+1)]}_{\rm CF}$. More generally, this relationship holds for $\Psi_B=\Psi^{[n_\uparrow, n_\downarrow]}_{\rm CF}$ and $\tilde{\Psi}_B=\Psi^{[-(n_\uparrow+1), -(n_\downarrow+1)]}_{\rm CF}$. The situation is similar to spinless CF states for fermions where a CF state with $n+1$ filled Lambda levels and reverse flux attachment is the particle-hole conjugate of the CF state with $n$ filled Lambda levels and direct flux attachment. 

The relation between bosonic spinful CF state with direct and reverse flux attachments through the particle-hole conjugation goes beyond Eq.~\eqref{eq:FillingShiftBosonParticleHoleBIQHE}. Indeed the $K$-matrices of the $\Psi^{[n,n]}_{\rm CF}$ and $\Psi^{[-n-1,-n-1]}_{\rm CF}$ states are particle-hole conjugate to each other. We illustrate this with the example of the $n=-2$ state at $\nu=4/3$ (the argument can be straightforwardly extended to general $n$). Take the $K$-matrix in Eq.~\eqref{eq:KMatrixNu43}, and redefine the last two components of the Chern-Simons gauge fields as $\tilde{a}_3=a_3+a_1+a_2$, $\tilde{a}_4=a_4+a_1+a_2$. The $K$-matrix then takes the form
\begin{equation}
K=\left(\begin{array}{cccc}
          -2 & -1 & 0 & 0 \\
          -1 & -2 & 0 & 0 \\
          0 & 0 & 0 & 1 \\
          0 & 0 & 1 & 0 \end{array}\right),
\end{equation}
with transformed charge vectors
\begin{equation}
	\tau_\up = \begin{pmatrix} 0 \\ -1 \\ 1 \\ 0 \end{pmatrix}, \quad
	\tau_\dn = \begin{pmatrix} -1 \\ 0 \\ 0 \\ 1 \end{pmatrix} .
\end{equation}
This is exactly the particle-hole transformed version of the Halperin $(221)$ state at $\nu=2/3$: the upper block of the $K$-matrix is a bare conjugate of the $(221)$ state, while the lower block is a bosonic integer quantum Hall state.

%%%%%%%%%%%%%%%%%%%%%%%%%%%%%%%%%%%%%%%%%%%%%%%%%%%%%%%%%%%%%%%%%%%%%%%%%%55
\section{Bosonic integer quantum Hall effect}\label{Sec:Nu2}

The physical properties of the bIQHE have been studied in Refs.~\onlinecite{Lu-PhysRevB.86.125119,Senthil-PhysRevLett.110.046801,Liu-PhysRevLett.110.067205,Geraedts2013288}. In particular, the Hall conductivity was shown~\cite{Lu-PhysRevB.86.125119} to be quantized and equal to an even integer. 
 There is reasonable numerical evidence of the bosonic IQHE in both continuous bilayer models~\cite{Furukawa-PhysRevLett.111.090401,Wu-PhysRevB.87.245123,Regnault-PhysRevB.88.161106} and lattice models~\cite{Moller:2009p184,Sterdyniak-PhysRevLett.115.116802,He-PhysRevLett.115.116803,Zeng-PhysRevB.93.195121}. The edge physics of the bIQHE consists of a charged chiral edge mode and a counter propagating neutral mode as given by the $K$ matrix of Eq.~\eqref{eq:KMatrixNu2}. Despite being non-chiral this edge structure is protected so long as charge conservation symmetry is preserved and is reflected in the entanglement spectrum of the bIQHE ground state~\cite{Furukawa-PhysRevLett.111.090401,He-PhysRevLett.115.116803}. 

\begin{figure}
\centerline{\includegraphics[width=0.92\linewidth]{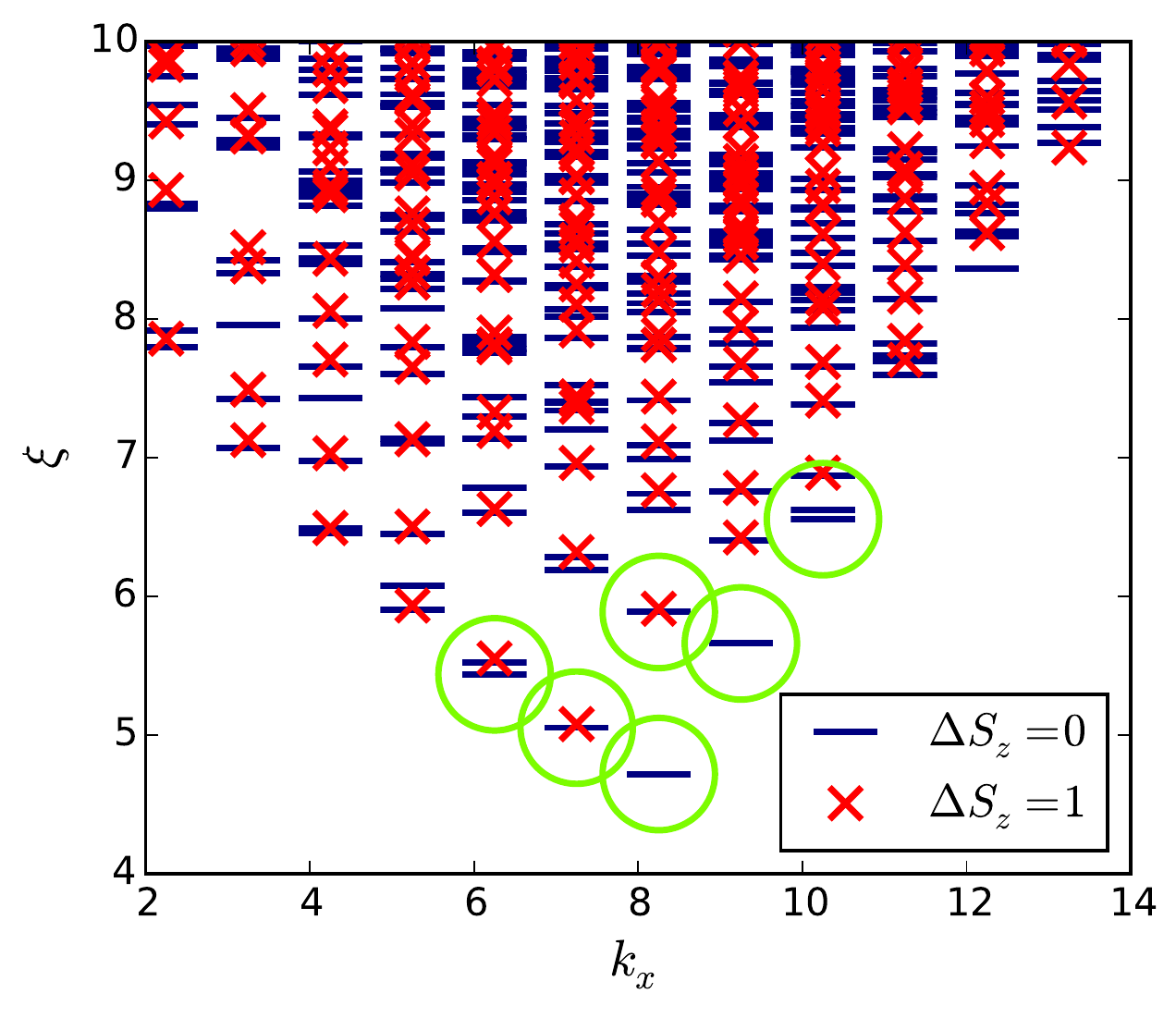}}
\caption{Entanglement spectra at $\nu=2$ as a function of the momentum along the cylinder perimeter $2 \pi k_x / L$ obtained from an iDMRG simulation at perimeter $L=16$ and bond dimension $5400$, using the charge sector with equal amounts of charge on either side of the cut (i.e., $\Delta N=0$) and the differences of spin $\Delta S_z=0$ and $\Delta S_z=1$. The spectrum at $\Delta S_z=0$ as the counting $1,1,2,...$ on both the left and right sides, indicative of both a left-moving a right-moving mode, and as expected for a bosonic IQHE. The additional states on the right side with $\Delta S_z=1$ are also consistent with this state. 
We circle the low-lying states which clearly match the predictions of the effective theory, at larger entanglement energy the states are too close together to determine whether they agree with predictions.
The slight loss of $SU(2)$ is a consequence of the bond dimension truncation.
 }
\label{fig:RSES_nu2}
\end{figure}

To our knowledge, the only iDMRG study of this phase was done on a lattice model~\cite{He-PhysRevLett.115.116803}. As a warm-up for our iDMRG approach of the continuous bilayer model setup, we have studied the emergence of such a phase at filling factor $\nu=2$. The larger system sizes accessible in iDMRG allow us to obtain an entanglement spectrum at $\nu=2$ less prone to finite size truncation, which has not been done before in a continuum model.  When plotting entanglement spectra in our spinful model, we must specify both the difference in charge between the left and right sides of the cut ($\Delta N$) and the difference in spin ($\Delta S_z$).
We plot the spectrum for $\Delta N=0$ and $\Delta S_z$ integer is shown in Fig.~\ref{fig:RSES_nu2}, and it has the predicted counting \cite{He-PhysRevLett.115.116803} that can be deduced from the $K$ matrix given in Eq.~\eqref{eq:KMatrixNu2} (see Appendix~\ref{app:KMatrixRSES}).
In particular, for a given charge sector the presence of one propagating and one counter-propagating mode implies that we should expect a counting of $1,1,2,...$ to both the left and the right in each charge sector.
Furthermore, the $K$ matrix can be used to determine the relationship between different charge sectors~\cite{Furukawa-PhysRevLett.111.090401,He-PhysRevLett.115.116803}. In brief, the momentum of the lowest-lying entanglement level in a given charge sector can be obtained from the formula:
\begin{equation}
k_0=\sum_i \frac{\lambda_i}{2} (\vec v_i \vec q)^2,
\label{k0}
\end{equation}
where $\lambda_i$, $\vec v_i$ are the eigenvalues and eigenvectors of the $K$-matrix and $\vec q$ is a vector representing the charge of an entanglement sector. More details can be found in Appendix~\ref{app:KMatrixRSES}. For the $K$ matrix of Eq.~\eqref{eq:KMatrixNu2} the above formula imples that the spectrum at $S_z=1$ should start at momentum one less than $S_z=0$, just as we observe.

In addition to the iDMRG study, we have also performed an exact diagonalization study on the sphere, considering slightly larger systems sizes than Refs.~\onlinecite{Furukawa-PhysRevLett.111.090401,Wu-PhysRevB.87.245123}. This study corroborates the emergence of the bIQHE at $\nu=2$ and we provide this information in the Appendix~\ref{app:overlaps}.

%%%%%%%%%%%%%%%%%%%%%%%%%%%%%%%%%%%%%%%%%%%%%%%%%%%%%%%%%%%%%%%

\section{Nature of the \texorpdfstring{$\nu=4/3$}{nu = 4/3} phase}\label{Sec:Nu43}

If an emergent bosonic particle-hole symmetry exists, we should expect that it relates FQH states with filling fractions $\nu$ and $2-\nu$. In this section we argue that the spinful bosonic states at $\nu=2/3$ and $\nu=4/3$ are indeed related by this symmetry. At $\nu=2/3$ a Halperin (221) state is the exact ground state of a Hamiltonian with only $V_0$ interaction between spins of different species and those of the same species. 

Under the emergent particle-hole symmetry, the Halperin $(221)$ state transforms into a $\Psi^{[-2,-2]}$ state as argued in Sec.~\ref{Sec:EmergentPHSymmetry}. We therefore want to find out whether the $\Psi^{[-2,-2]}_{\rm CF}$ is the ground state at $\nu=4/3$. Its main competitor is the $k=2$ NASS state mentioned in Sec.~\ref{Sec:ModelAndCFStates}. Previous exact diagonalization studies could not access large enough system sizes to definitively rule out either candidate, though they found slightly larger overlaps for the $\Psi^{[-2,-2]}$ state~\cite{Wu-PhysRevB.87.245123} but also potentially the NASS topological degeneracy on the torus~\cite{Furukawa-PhysRevA.86.031604}. We have used iDMRG methods to determine that $\Psi^{[-2,-2]}$ is indeed the ground state. This determination comes from two pieces of evidence: the shift and the entanglement spectrum.\cite{Tu-PhysRevB.88.195412,Mong2017ToAppearSoon}

\begin{figure}
\centerline{\includegraphics[width=0.92\linewidth]{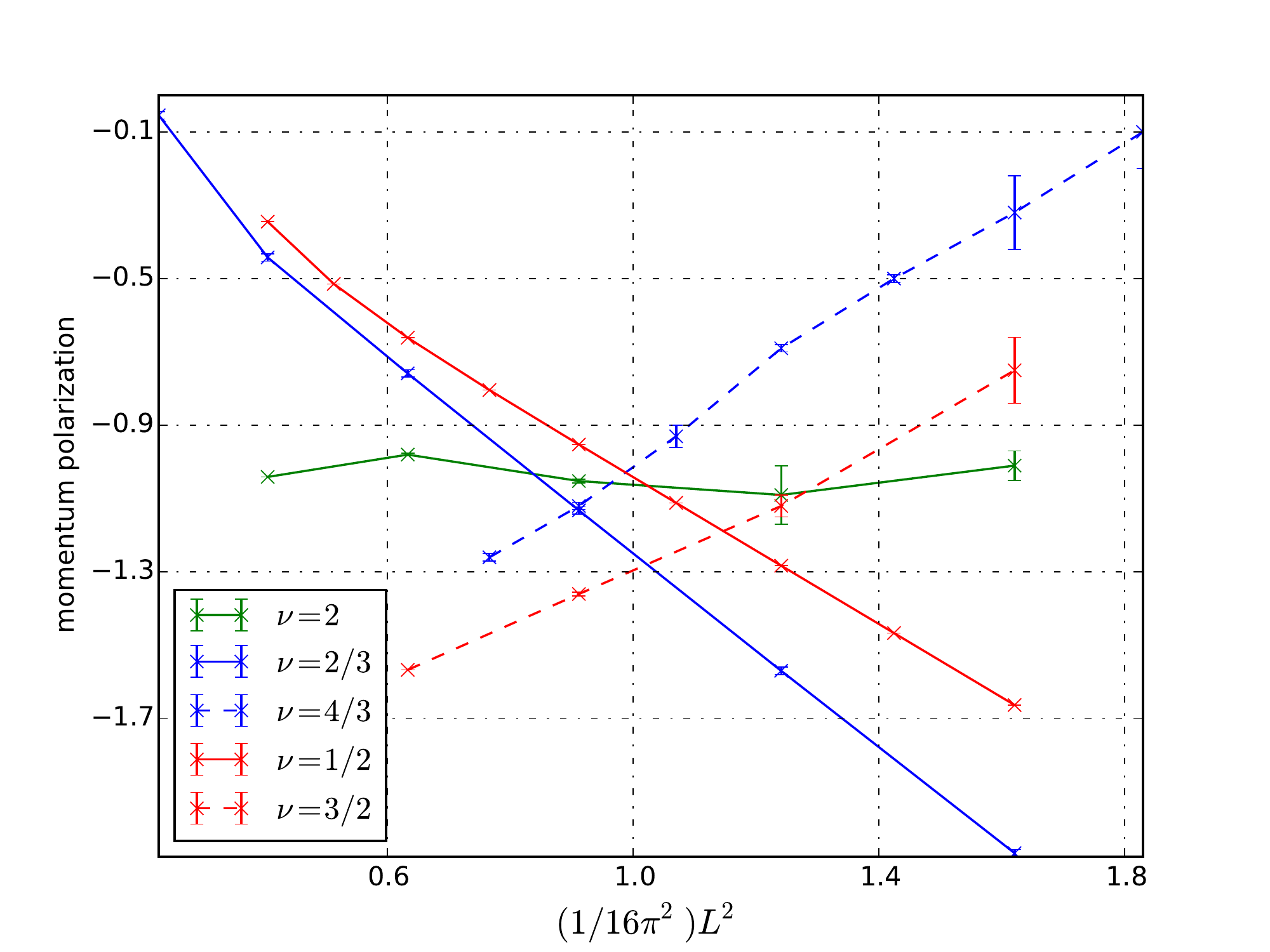}}
\caption{Momentum polarization for a number of system sizes and filling fractions, obtained using iDMRG. When plotted against $L^2$ as suggested by Eq.~\eqref{eq:MomentumPolarization}, the slope of the data gives $-\nu\delta$. This allows us to check Eq.~(\ref{eq:FillingShiftBosonParticleHoleBIQHE}). We see that the equation is satisfied for the particle-hole conjugate pairs $(2/3,4/3)$ and $(1/2,3/2)$. Further we see that the shift at $\nu=2$ is $0$, as expected. }
\label{fig:mompol}
\end{figure}

While we consider an infinite cylinder, the shift $\delta$ can be computed~\cite{Zaletel-PhysRevLett.110.236801} from the momentum polarization~\cite{Tu-PhysRevB.88.195412, Mong2017ToAppearSoon}, which has the following dependence on cylinder circumference $L$ (we set the magnetic length to one):
\begin{equation}
{\rm momentum~polarization}=-\delta\frac{\nu}{16\pi^2} L^2+O(1).\label{eq:MomentumPolarization}
\end{equation} 
We plot the momentum polarization vs. $L^2$ for a number of different spinful bosonic cases in Fig.~\ref{fig:mompol}. By dividing the slope of such data by the filling fraction, we obtain the expected values of $\delta=0,2$ for $\nu=2,2/3$ respectively. For $\nu=4/3$, we obtain $\delta=-1$, consistent with the $\Psi^{[-2,-2]}$ state but not the NASS state, which has shift $2$.

\begin{figure}
\centerline{\includegraphics[width=0.92\linewidth]{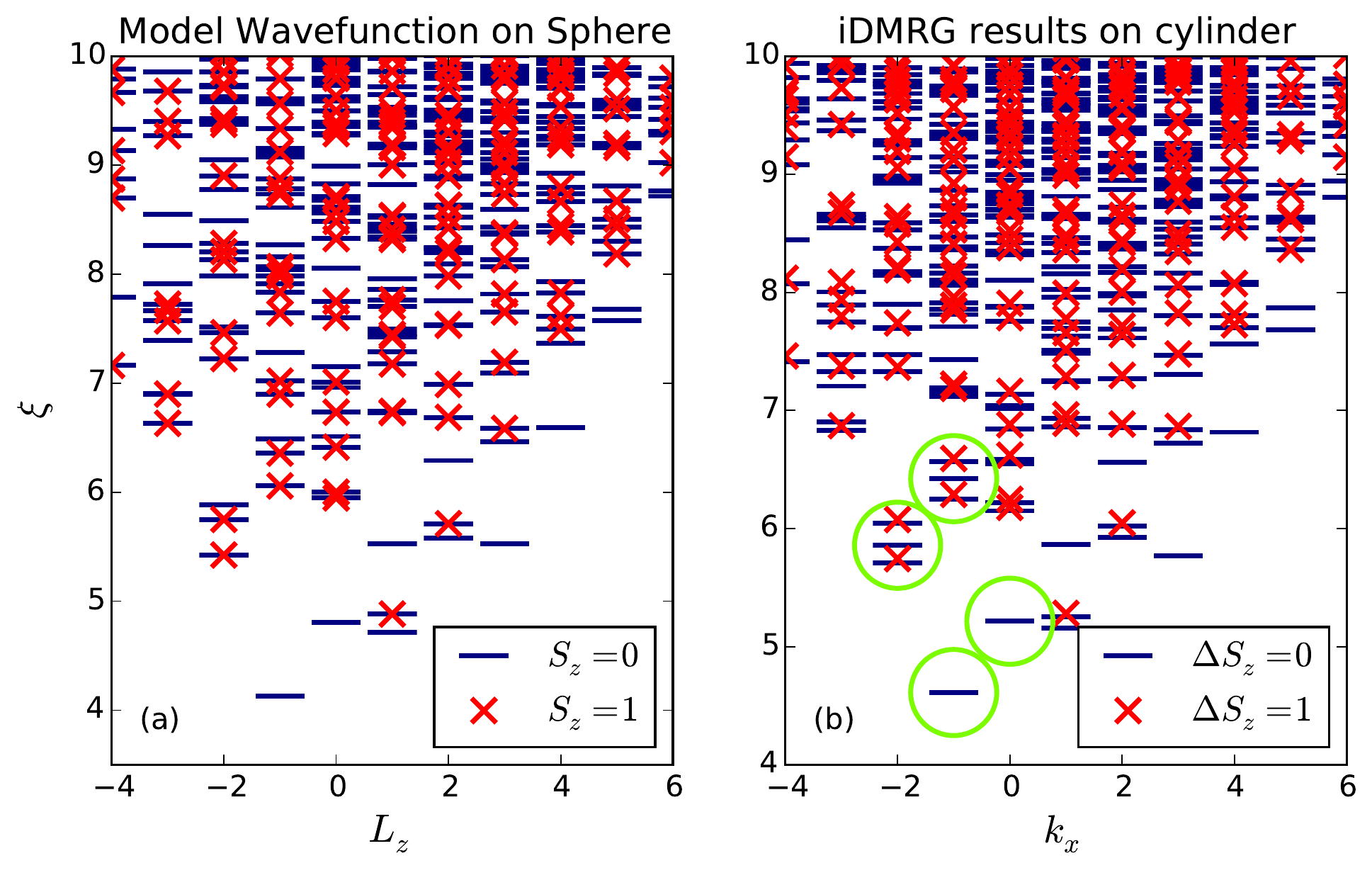}}
\caption{Entanglement spectra at $\nu=4/3$ (a) from a model wavefunction on a sphere with $N=16$, $N_A=8$, and (b) from an iDMRG simulation at $L=17$, $\Delta N=0$, and bond dimension $5400$. In both cases there is an equal amount of charge on either side of the entanglement cut. Both spectra have the same counting a low entanglement energies, a counting consistent with the $\jain$ state. Despite being on different geometries which could affect the shape of entanglement spectrum, we find a remarkable agreement between the model state on a finite sphere and the iDMRG on an infinite cylinder with the hardcore interaction.}
\label{fig:RSES_nu43}
\end{figure}

In Fig.~\ref{fig:RSES_nu43} we show the real-space entanglement spectrum for the $\jain$ state, obtained both from the model wavefunction on a sphere for $N=16$ bosons (a) and from iDMRG on an infinite cylinder (b). 
For the model wavefunction we specify for one subsystem the charge sector by $N_A$ and the total number of bosons, and $S_{z,A}$, the total spin. Compared to specifying the $\Delta N$ as in the iDMRG, this approach shifts all momenta by the same amount but otherwise does not affect our analysis.
The two spectra exhibit the same counting at low entanglement energies, where they also agree with the predictions from the edge theory. Similarly to the bIQHE, the counting can be deduced from the $K$ matrix given by Eq.~\eqref{eq:KMatrixNu43}. In particular, we expect a counting of $1,1,2,...$ to the right and $1,3,9,...$ to the left, and we expect the spectrum at $\Delta S_z=1$ to be two-fold degenerate and start at momentum one less that the $\Delta S_z=0$ data. We find that the low-lying levels in Fig.~\ref{fig:RSES_nu43} indeed match these predictions.

%%%%%%%%%%%%%%%%%%%%%%%%%%%%%%%%%%%%%%%%%%%%%%%%%%%%%%%%%%%%%%%
\section{Particle-hole symmetry between \texorpdfstring{$\nu=1/2$ and $\nu=3/2$}{fillings 1/2 and 3/2}}\label{Sec:Nu32}

The states so far considered in this work are all spin singlets. We can test whether the bosonic particle-hole symmetry also applies to states beyond this specific class. In this section, we consider the particle-hole conjugate of a state at filling fraction $\nu=1/2$, where the electrons are all constrained to be spin-polarized. For the hardcore interaction, the Laughlin state is the exact and unique fully polarized state at this filling factor. It can be written in the CF state language as $\wf{1}{0}$. A natural candidate proposed in Sec.~\ref{Sec:EmergentPHSymmetry} for the particle hole conjugate of the $\nu=1/2$ Laughlin state is the $\wf{-2}{-1}$. Note that this state is partially spin polarized (exactly as expected).  It was also discussed in Ref.~\onlinecite{Wu-PhysRevB.87.245123}.

\begin{figure}
\centerline{\includegraphics[width=0.92\linewidth]{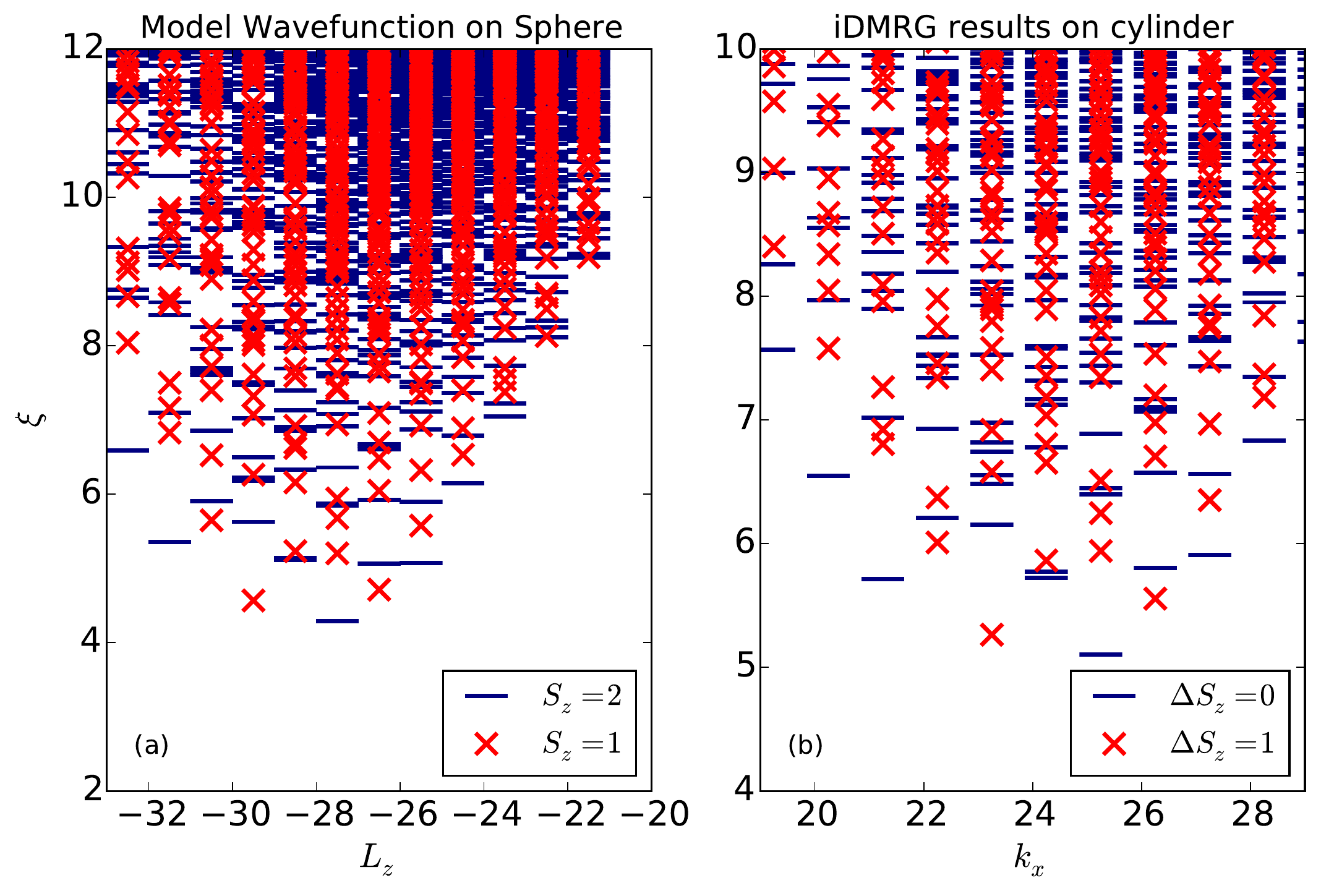}}
\caption{Entanglement spectra at $\nu=3/2$ (a) from a model wavefunction on a sphere with $N=17, N_A=8$, and (b) from an iDMRG simulation at $L=16, \Delta N=0$ and bond dimension $5400$. Both spectra have the same counting a low entanglement energies, a counting consistent with the $\wf{-1}{-2}$ state.  }
\label{fig:RSES_nu32}
\end{figure}

As opposed to the situation at $\nu=4/3$ described in Sec.~\ref{Sec:Nu43}, $\wf{-2}{-1}$ does not describe the absolute ground state of the hardcore interaction at $\nu=3/2$, only the corresponding polarization sector. This could still be relevant when adding a polarization (Zeeman) field (the Zeeman field is odd under particle-hole transform). Thus we can search for evidence of this state using similar methods to the previous section. We performed iDMRG calculations at $\nu=1/2$ and $\nu=3/2$, fixing the fillings in the individual components to the above-mentioned values. In Fig.~\ref{fig:mompol} we have shown the resulting momentum polarization for $\nu=1/2$ and $\nu=3/2$, and they indeed satisfy Eq.~\eqref{eq:FillingShiftBosonParticleHoleBIQHE}. 
Furthermore in Fig.~\ref{fig:RSES_nu32} we show the real-space entanglement spectra for the $\nu=3/2$ case, with the spectrum of the model wavefunction on a sphere in (a) and the iDMRG results in (b). We see that the low-lying part of the spectra are very similar. The form of the spectra can be determined from Eq.~\eqref{k0}, however this analysis is complicated since the $K$ matrix (here 3) is larger than the number of conserved quantities (here 2, namely $\Delta N$ and $\Delta S_z$).
\footnote{This is also a problem at $\nu=4/3$, but in that case the lowest-lying states are not affected.}.
We provide in Appendix~\ref{app:KMatrixRSES} an extensive discussion that shows that  real-space entanglement spectrum does indeed follow from the $K$ matrix associated with the $\wf{-2}{-1}$ CF state. 

We can wonder if slightly tuning the interaction could partially polarize the system ground state at this specific filling factor. For example, we can add some $V_1$ pseudo-potential as an additional knob while preserving the $SU(2)$ symmetry. For $\nu=1$~\cite{Liu-PhysRevB.93.085115} and $\nu=4/3$ (see Appendix~\ref{app:overlaps}), the previous description holds true for $V_1 \lesssim 0.3$ beyond which the system spontaneously fully polarizes. From our exact diagonalization on both the sphere and the torus geometries, the only option for a ground state that has the same polarization as $\wf{-2}{-1}$ would be a narrow region close to the transition toward the fully polarized regime.

%%%%%%%%%%%%%%%%%%%%%%%%%%%%%%%%%%%%%%%%%%%%%%%%%%%%%%%%%%%%%%%
\section{Evidence of the emergent CF Fermi sea at \texorpdfstring{$\nu=1$}{nu = 1}}\label{Sec:Nu1}

The phase diagram of interacting bosons at filling factor $\nu=1$ hosts a large variety of phases~\cite{Read-PhysRevB.58.16262, Liu-PhysRevB.93.085115} depending on the interaction. For spinless bosons, previous studies~\cite{Cooper-PhysRevLett.87.120405,regnault-PhysRevLett.91.030402,Chang-PhysRevA.72.013611,Regnault-PhysRevB.69.235309,Regnault-PhysRevB.76.235324,Liu-PhysRevB.93.085115,Wu-PhysRevB.87.245123} have shown strong evidence that the two-body hardcore interaction leads to an emerging Moore-Read state. Among the other possible phases are two decoupled copies of the Laughlin $\nu=1/2$ state (i.e., the Halperin $(220)$ state) or the coupled Moore-Read state~\cite{Hormozi-PhysRevLett.108.256809}. Restricting to the pure $SU(2)$ symmetric hardcore interaction, Ref.~\onlinecite{WuCFL} has provided hints using exact diagonalization on the sphere geometry of a possible CFL emergence. 

Finding a particle-hole symmetric state at $\nu=1$ would provide further evidence that an emergent particle-hole symmetry exists.
A CFL naturally exhibits a particle-hole symmetry if the CF are Dirac fermions~\cite{Son-PhysRevX.5.031027,Wang-PhysRevB.93.085110,Geraedts197}, while the particle-hole symmetric nature of the CF Fermi sea if the CF are non-relativistic fermions (or equivalently the Halperin-Lee-Read~\cite{Halperin-PhysRevB.47.7312}) has recently raised opposite views \cite{Barkeshli-PhysRevB.92.165125,LevinSon,Wang-2017arXiv170100007W}. It was also argued in Ref.~\onlinecite{Wang-PhysRevB.94.245107} that for two-component bosons at $\nu=1$, if both the $SU(2)$ symmetry and particle-hole symmetry are preserved, the system cannot be gapped (even with topological order). Therefore a CFL state at $\nu=1$ is highly anticipated if particle-hole symmetry indeed holds.

To numerically study in an unbiased way the physics at $\nu=1$, we can rely on either the iDMRG or finite size exact diagonalizations on the torus geometry. We first present results obtained by exact diagonalization in the torus geometry.
 We have computed the low energy spectrum of the $\nu = 1$ system for the hardcore interaction with up to $N = 14$ bosons. In order to experiment with different discrete symmetry groups, we adjust the angle $\theta$ between the spanning vectors of the torus. We choose $\theta = \pi / 3$ and $\theta = \pi / 2$ (square torus) to obtain the $C_{6v}$ and $C_{4v}$ symmetries, respectively. 

\begin{figure}[htb]
\includegraphics[width=0.95\linewidth]{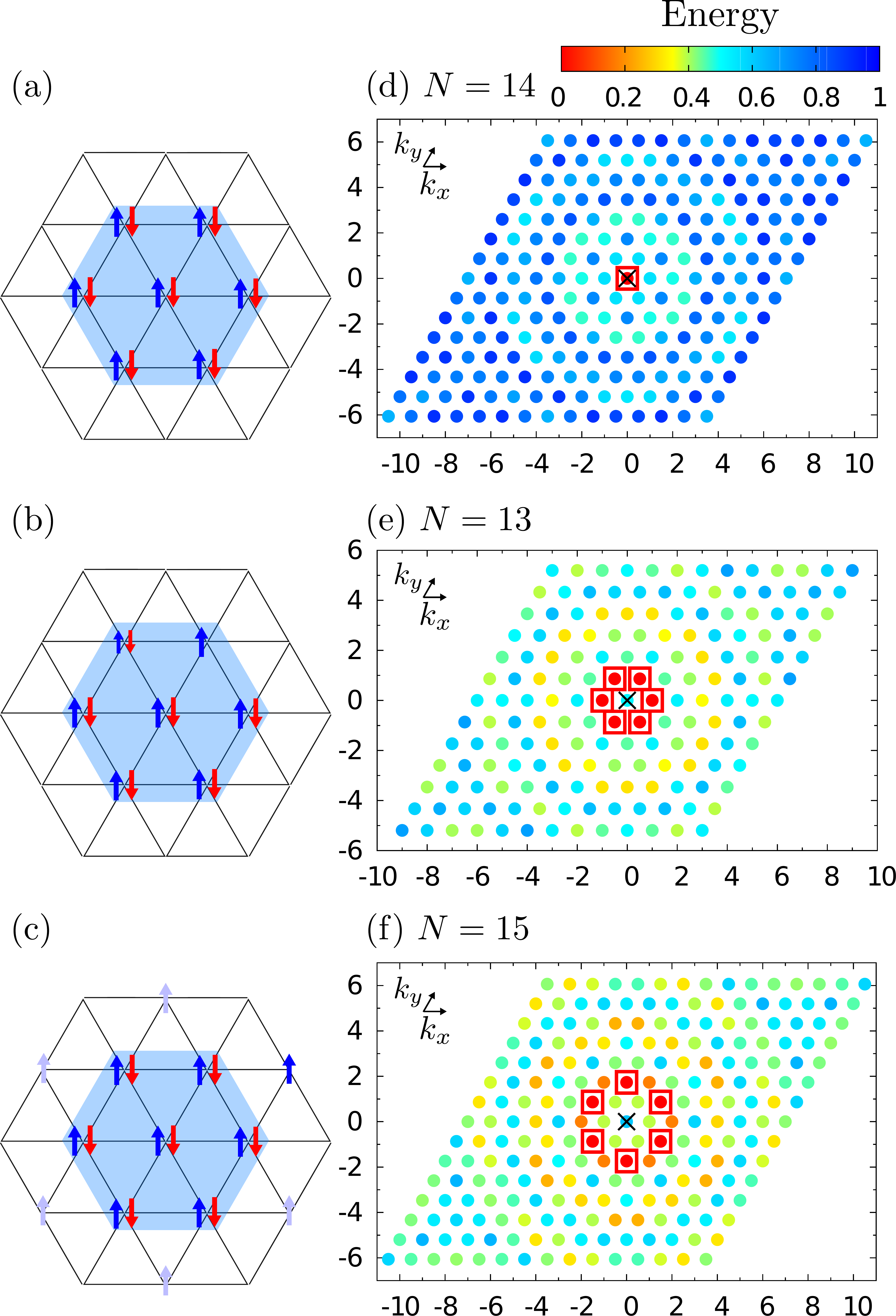}
\caption{{\it Left panel}: Filling up the Brillouin zone with free spinful composite fermions to form a Fermi sea on a torus with twisting angle $\theta = \pi/3$ (a, b, c). This picture predicts a unique ground state for $N=14$ (a). Removing one composite fermion on the outer shell of the Fermi sea creates a quasihole excitation (b). Adding a composite fermion to one the the sites with the shortest distance to the center of the Fermi sea creates a quasielectron excitation (c). For the quasielectron excitation we depict the equivalent positions in reciprocal space in light blue. {\it Right panel}: Lowest energies in each momentum sector for the hardcore hamiltonian at $N=N_\Phi$ on a torus with twisting angle $\theta = \pi/3$ (d, e, f). The lowest energies are indicated by a red box. (d), (e) and (f) involve respectively $N=14$, $13$ and $15$. Plots (e) and (f) are centered (black cross) around $(0,0)$ and plot (d) is centered around $(\pi,\pi)$.}
\label{fig:hexagonCFFS}
\end{figure}

Considering the composite fermions as free particles, we can predict the degeneracy and momentum sectors of the ground state for some given numbers of particles (see Fig.~\ref{fig:hexagonCFFS}). This picture provides a description of the ground state at some commensurable sizes as well as its charged quasiparticle or quasihole excitations. The same description can be applied to spinless fermions at $\nu = 1/2$ and it predicts the degeneracy and momentum sectors of the ground state, up to an overall $(\pi, \pi)$ shift~\cite{geraedts-inpreparation}. This predictive description in terms of free CF, as well as the analogy with the fermionic case where a CFL is expected are strong arguments in favor of a bosonic CFL.

Charged excitations provide a very crisp illustration of the finite size CF Fermi sea construction. For $\theta = \pi/3$, the $C_{6v}$ symmetry imposes a unique ground state when, for instance, the number of spinful CF is $N=14$ (see Fig.~\ref{fig:hexagonCFFS}a). Removing one composite fermion will generate a quasihole state (see Fig.~\ref{fig:hexagonCFFS}b), which is sixfold degenerate. Similarly, adding one composite fermion will create a quasiparticle excitation with degeneracy $6$ in the $S = 1/2$ sector (see Fig.~\ref{fig:hexagonCFFS}c). Our exact diagonalization data supports this image as shown in Figs.~\ref{fig:hexagonCFFS}d, e and f. The position of the origin in reciprocal space depends on the parity of the number of particles (like for spinless fermionic systems): the singlet ground state at $N = 14$ lies in momentum sector $(\pi, \pi)$, while the $6$ states at $N = 13$ are centered around $(0,0)$. In principle, we could apply a similar approach for the low energy neutral excitations but there finite size effects are more important and remain to be understood. Similiar results for the square torus are given in Appendix~\ref{app:CFLEvidence}.

\begin{figure}
\includegraphics[width=\linewidth]{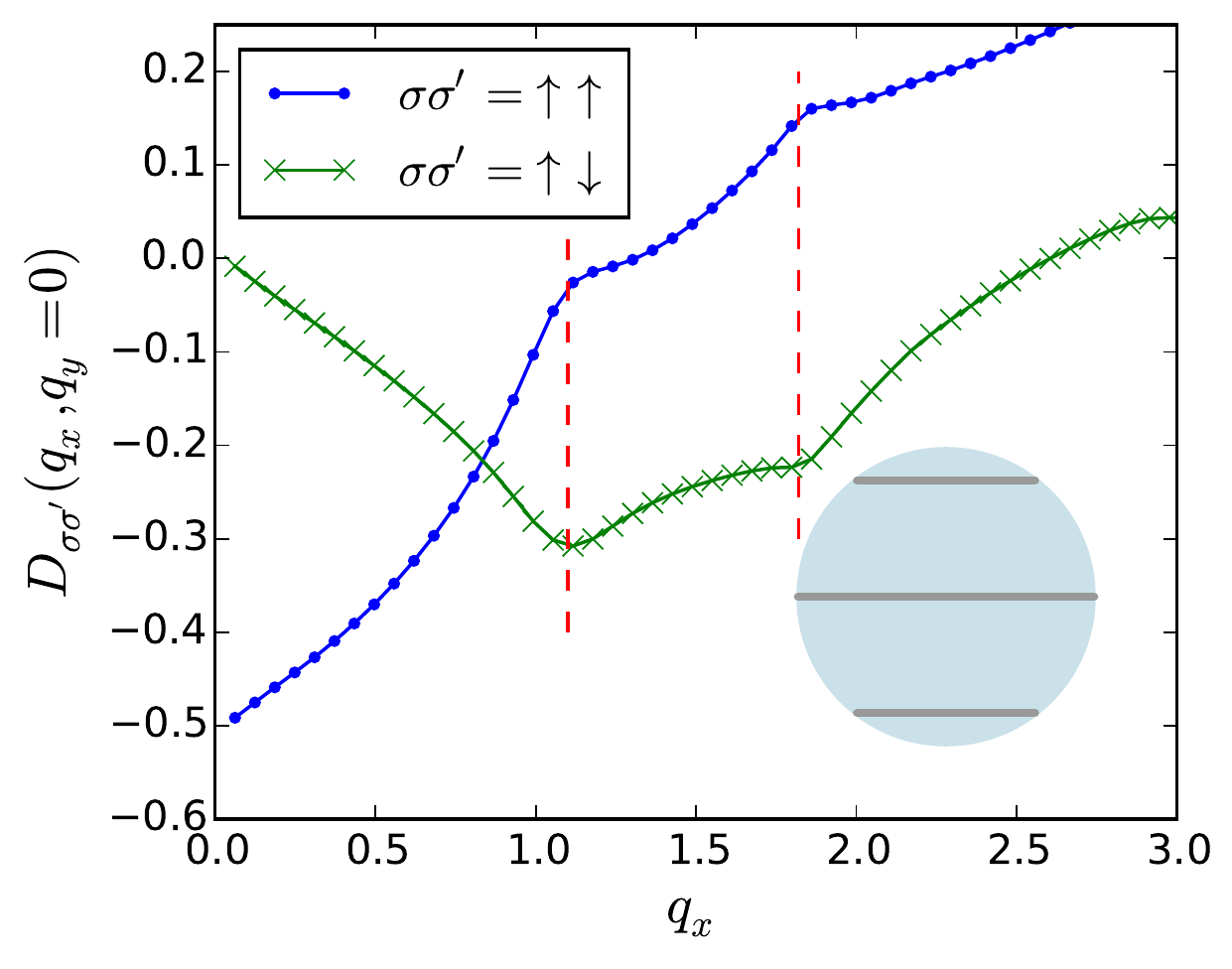}
\caption{Guiding center structure factor for a system with $L=8$, with $q_y=0$ as defined in Eq.~\eqref{eq:Dq}. In our system with periodic BC in the $y$ direction, only discrete value of $q_y$ are possible. We remind that $D_{\up\up}=D_{\dn\dn}$. The singularities in the figure correspond to the size of the Fermi surface at different values of $q_y$. The inset shows an example of a circular Fermi surface with two `wires', the lengths of the wires are given by the location of the singularities.}
\label{structure}
\end{figure}

We can apply iDMRG analysis similar to that of Ref.~\onlinecite{Geraedts197} to the bosonic CFL at $\nu=1$. 
Our method is to search for singularities in the momentum-space guiding center structure factor:
\begin{align}
	D_{\sigma\sigma^\prime}(\vec q) = \braket{\norder{ \rho_\sigma(\vec q)\rho_{\sigma^\prime}(-\vec q) }} e^{|\vec q|^2/2}.
	\label{eq:Dq}
\end{align}
This quantity has a singularity whenever $\vec q$ corresponds to a process which hops a composite fermion from one part of the Fermi surface to another. 
The indices $\sigma$, $\sigma^\prime$ represent spin species. An example of such data, for $L=8$, is shown in Fig.~\ref{structure}.
Since we work on a cylinder of finite radius in the $y$ direction, only certain discrete values of $q_y$ are allowed. By fixing $q_y$ and measuring the $q_x$ at which singularities occur, we can map out the composite fermion Fermi surfaces. 
We have found that $D_{\up\up}$ and $D_{\dn\dn}$ are identical while $D_{\up\dn}$ has singularities in the same locations, implying that there is an identical Fermi surface in both layers and consistent with the prediction that the CFL is a spin singlet. Elsewhere in this work we used a $V_0$ pseudo-potential interaction, but we find that with such an interaction the DMRG does not converge, therefore for our DMRG study at $\nu=1$ our bosons interact via a Coulomb repulsion. Shorter range interactions always induce more finite size effect for the CFL\cite{WuCFL}, including for spinless fermions\cite{Regnault-PhysRevB.70.241307}. Thus using the Coulomb interaction is merely a trick to improve the convergence rather than a drive to another phase.

\begin{figure}[htb]
\includegraphics[width=\linewidth]{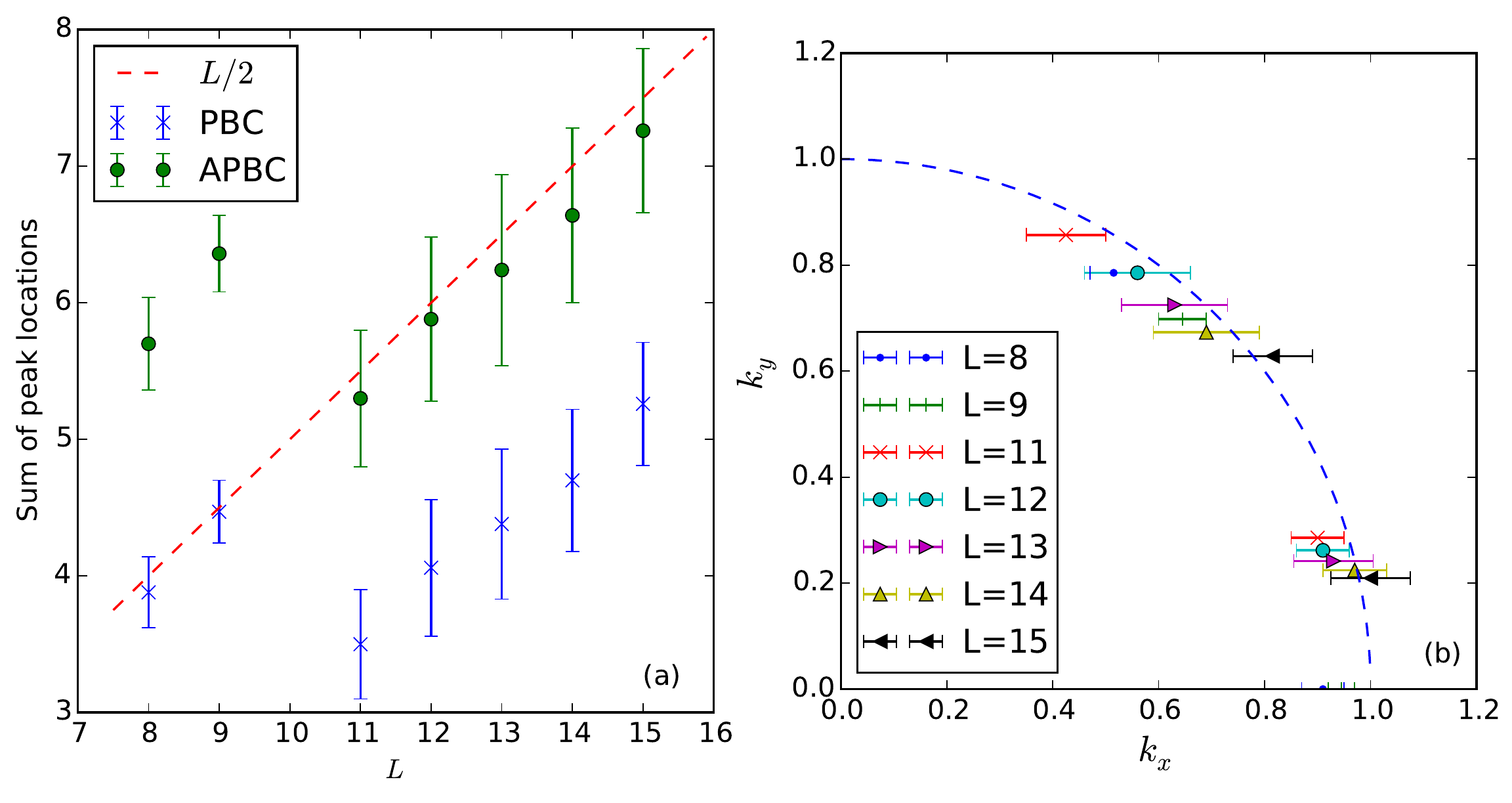}
\caption{Using the singularites in Fig.~\ref{structure} to determine the shape of the Fermi surface. In (a) we determine whether the composite fermions have periodic boundary conditions (PBC) or antiperiodic boundary conditions (APBC), by plotting the sums of the singularities assuming both cases, and seeing which sum matches Luttinger's theorem, which constrains this sum to be equal to the electron density, which is $L/2$. 
In (b) we use the appropriate boundary conditions to plot all the locations of the singularities in momentum space. We find good agreement between the locations of the singularites and the expected Fermi surface with $k_F=1$.}
\label{CFLDMRG}
\end{figure}

On a cylinder geometry, the composite fermions do not need to have the same boundary conditions as the microscopic degrees of freedom, and therefore in order to  map the Fermi surface we need to determine which boundary conditions are present. We can do this by appealing to Luttinger's theorem, which implies that the lengths of the `wires' in the inset of Fig.~\ref{structure} must add up to the total electron density. If we have two identical Fermi surfaces at $\nu=1$, this density is $L/2$. Only one set of boundary conditions (BC) can satisfy Luttinger's theorem, since if we have periodic BC the longest wire appears once at $k_y=0$, while if we have antiperiodic BC it appears twice at $k_y=\pm \pi/L$. We plot the sums of the wire lengths for a number of system sizes in Fig. ~\ref{CFLDMRG}(a), assuming both periodic and antiperiodic BC. We see that the data for $L=8,9$ obey periodic BC, while at $L=11-15$ we have antiperiodic BC
\footnote{Which BC are preferred is a question of energetics, though from the fermionic case \cite{Geraedts197} we expect it to be periodic in $L$ as the system tries to avoid having a wire near the edge of the Fermi surface.}.

Once the boundary conditions have been determined we can plot the locations of the singularities and compare them to the expected circular Fermi surface with $k_F=1$.  The $q_x$ of these singularities are determined from data such as Fig.~\ref{structure} while the $q_y$ are determined from the boundary conditions. The data is plotted in Fig.~\ref{CFLDMRG}(b), where we see good agreement with the expected circle. Deviations from a perfect circle are finite-size effects related to the need to satisfy Luttinger's theorem. 

The error bars on the DMRG data for the CFL at $\nu=1$ are larger than those in Ref.~\onlinecite{Geraedts197} due to the higher computational cost of simulating a two-component bosonic system instead of a single-component fermionic one. Additional iDMRG results are discussed in Appendix~\ref{app:CFLEvidence}.

\section{Conclusion}\label{Sec:Conclusion}

The effects of particle-hole symmetry in the lowest Landau level are a subject of much activity, both historically for Laughlin states as well as recently for the composite Fermi liquid and various non-Abelian states at $\nu=1/2$. 
In this article, we have provided numerical evidence that a particle-hole symmetry is emergent for spinful bosons in the lowest Landau level. By using a modern numerical techniques including exact diagonalization and iDMRG, we were able to show that the low energy physics at $\nu=4/3$ is related to the Halperin $(221)$ state by the particle-hole symmetry, settling once and for all the nature of the phase at $\nu=4/3$. This symmetry also extends to non-spin singlet states such as the $\nu=3/2$ partner of Laughlin $\nu=1/2$ state. At the particle-hole symmetric invariant filling factor $\nu=1$, we find evidence for a composite Fermi liquid which has a Fermi surface with $k_F=1$ in each spin component. 

A natural question is how robust is this symmetry. While in the fermionic case the symmetry is exact for all  two-body interactions, for bosons the symmetry applies only at low energies and is not guaranteed to work for any interaction. Nonetheless we find that slightly modifying the hardcore interaction by including additional pseudo-potential does not affect our results significantly (see Appendix~\ref{app:overlaps}), suggesting that it might be stable to a range of two-body interactions. The validity of our finding to non-singlet states suggests that breaking the $SU(2)$ would not necessarily lead this symmetry to disappear. However, we have found no signature of a particle-hole symmetry if we strictly focus on the fully polarized sector, i.e., for single component bosons. We know that the particle-hole symmetry is not present at $\nu=1$ since the Moore-Read state which breaks it, is a valid description of the low energy physics at $\nu=1$. We also have observed some signatures of the Read-Rezayi state at single-component $\nu=3/2$, which is clearly not the particle-hole conjugate of the Laughlin state at $\nu=1/2$. Moreover we did not find any clear evidence of a bIQHE at $\nu=2$ (it can emerge on a lattice model with strictly hardcore interaction \cite{He-2017arXiv170300430H}). All these are consistent with the intuition that the emergence of particle-hole symmetry is much more natural if a composite fermi liquid phase is stabilized at $\nu=1$.

Now that the particle-hole symmetry for bosons has been established, a number of the current questions about particle-hole symmetry in the fermionic case can also be asked of the bosonic one. 
A microscopic understanding of the emergent particle-hole symmetry would be helpful to such analysis, especially for addressing the role of particle-hole symmetry in the bosonic CFL. It would also pave the way to finding interactions whose low energy physics is described by phases such as the anti-Pfaffian or particle-hole symmetry Pfaffian. 
A candidate theory\cite{Wang-PhysRevB.94.245107} for a CFL with emergent particle-hole symmetry in this system has  two species of Dirac composite fermions  at finite density, and the associated Fermi surface Berry phase of $\pi$. Demonstrating this numerically is an interesting future challenge.
These interesting problems will be developed in future works. 

\begin{acknowledgments}
We thank M.~Zaletel, M.~Hermanns, E.~Ardonne, B.~Bradlyn, J.~Cano and Y.~C.~He for fruitful discussions. N.~R. was supported by Grant No.~ANR-16-CE30-0025. S.~G. was supported by Department of Energy  BES Grant DE-SC0002140. C.~W. was supported by the Harvard Society of Fellows. T.S.\   is supported  by a  US Department of Energy grant DE-SC0008739. T.S.\  was also partially supported by a Simons Investigator award from the Simons Foundation. 
\end{acknowledgments}

\bibliography{bosons_emergent_ph}

\begin{appendix}

\section{\texorpdfstring{$K$}{K}-matrices of composite fermion states}
\label{KmatrixAppendix}

In this Appendix we derive the $K$-matrix for the Jain state $\Psi_{\rm CF}^{[n,n]}$. The result can be extended straightforwardly to any $\Psi_{\rm CF}^{[n_{\uparrow},n_{\downarrow}]}$.

Consider a two-component bosonic system, in a Jain state where the composite fermions $\psi_{\uparrow,\downarrow}$ fill $2n$ Landau levels. At the level of effective field theory, before we integrate out the composite fermions, the system should be described by the following effective Lagrangian:
\begin{equation}
\label{action}
\mathcal{L}=\mathcal{L}_0[\psi_{\up},\psi_{\downarrow},a_{\mu}]-\frac{1}{2\pi}b\; da-\frac{1}{4\pi}b\; db,
\end{equation}
where $a,b$ are emergent $U(1)$ gauge fields. Notice that $b$ has a self Chern-Simons (CS) term at level $-1$, which is trivial from topological quantum field theory (TQFT) point of view. So one can integrate it out and leave $a$ as the only emergent gauge field, with a self CS term at level $+1$. This is the usual form of action seen in the literature. However, one should be careful about the chiral central charge: the level-$1$ CS term is almost trivial except for its contribution to the chiral central charge. Since we care about chiral central charge, let's keep $b$ for now.

Now we integrate out $\Psi$ fermions in Eq.~\eqref{action}, keeping in mind that each occupied Landau level $\Lambda_i$ introduces an emergent gauge field $a_i$ with CS level $(-1)$ that couples with $a$ through $-\frac{1}{2\pi}a\; da_i$.  Also notice the usual definition of $K$-matrix has an additional minus sign through $\mathcal{L}=-\frac{1}{4\pi}a\;K\; da$. We then get a $(2n+2)\times(2n+2)$-dimensional $K$-matrix
\begin{equation}
\label{Knf}
\tilde{K}^{[n,n]}=\left(\begin{array}{cccccc}
          0 & 1 & 1&  1 & \cdots & 1 \\
          1 & 1 & 0 & 0 & \cdots & 0 \\
          1 & 0 & \operatorname{sgn}(n) & 0 & \cdots & 0  \\
          1 & 0 & 0 & \operatorname{sgn}(n) & \cdots & 0  \\
          \vdots & \vdots & \vdots & \vdots & \ddots & \vdots \\
          1 & 0 &  0 & 0 & \cdots & \operatorname{sgn}(n)   \end{array}\right),
\end{equation}
where the first column represents the gauge field $a$, the second column represents $b$, and the rest represent $a_i$ ($1\leq i\leq 2n$). 

However, the above $\tilde{K}$-matrix is not quite ready for immediate use (say, for edge states). This is because the first component (gauge field $a$ in Eq.~\eqref{action}) is not an ordinary $U(1)$ gauge field: it couples only to fermions (more precisely, fields that carry odd gauge charge are fermionic) rather than scalar bosons as in usual $K$-matrix theory. This means that one cannot directly use it to get the edge Luttinger liquid: for ordinary gauge field, the ``vacuum" outside of the system can be thought of as a condensate of scalar charges -- but this will not be an option if the only charge-$1$ field is fermionic. In formal term this kind of $U(1)$ gauge field is called spin$_c$ connection. Therefore it is more convenient to integrate out this spin$_c$ connection. Here this is possible because the term $\frac{1}{2\pi}a\;d(b+\sum_{i=1}^{2n}a_i)$ is a trivial TQFT, in which the $a$ gauge field serves merely as a Lagrange multiplier. Integrating out $a$ simply sets $b+\sum_{i=1}^{2n}a_i=0$. Now substituting $b=-\sum_{i=1}^{2n}a_i$ back into the $\tilde{K}$-matrix gives exactly the result in Eq.~\eqref{Kn}.

%%%%%%%%%%%%%%%%%%%%%%%%%%%%%%%%%%%%%%%%%%%%%%%%%%%%%%%%%%%%%%%%%%%%%%%%%%%55
\section{\texorpdfstring{$K$}{K} matrix and real-space entanglement spectrum}\label{app:KMatrixRSES}

The relation between the $K$ matrix and the real-space entanglement spectrum was previously discussed for the bIQHE at $\nu=2$ in Refs.~\onlinecite{Furukawa-PhysRevLett.111.090401, He-PhysRevLett.115.116803}.  When a model wavefunction contains multiple Lambda levels, then the dimension of the $K$ matrix is larger than the number of conserved quantities that one can specify numerically.
Very few studies have been performed relating the edge structure and the entanglement spectra~\cite{Regnault-PhysRevLett.103.016801,Rodriguez-PhysRevB.88.155307} in this case, and those that exist are limited to two Lambda levels and direct flux attachment. In this Appendix we exemplify the connection between the $K$ matrix and real-space entanglement spectrum, when we have two  Lambda levels {\it and} reverse flux attachment. We will focus on the case of the fraction $\nu=3/2$ of Sec.~\ref{Sec:Nu32} and the CF state $\wf{-2}{-1}$. Its $K$ matrix is given by

 \begin{equation}
 \label{eq:KMatrixNu32}
K^{[-2,-1]}=\left(\begin{array}{ccc}
          0 & 1 & 1 \\
          1 & 0 & 1 \\
          1 & 1 & 0 \end{array}\right),
\end{equation}

The rows (or columns) of $K^{[-2,-1]}$ are related to the variation of the CF number per Lambda level and per spin component that we denote $\Delta N_{1,\uparrow}$ and $\Delta N_{1,\downarrow}$ for the lowest Lambda level and $\Delta N_{2,\uparrow}$ for the second Lambda level. 
$\Delta N_i$ represents the difference in the number of a type of bosons on, say, the left side of an entanglement cut to the number of bosons that would be in that region if bosons were distributed uniformly. Such a definition is necessary in iDMRG, where total number of bosons is infinite, but in the exact diagonalization data on a finite system we can replace $\Delta N_i$ by $N_i$, which is just the total number of bosons in a region. The definitions are equivalent (up to an overall shift in the momentum of all entanglement levels).
The $K$ matrix indicates that we have one propagating mode with eigenvector $2$ and associated $U(1)$ charge $\Delta_Q$. Up to a normalization constant, $\Delta_Q$ is precisely the $\vec v_0 \vec q$ of Eq.~\eqref{k0}. We also find two counter-propagating modes with eigenvalues $-1$ and associated $U(1)$ charges ($\vec v_i \vec q$) $\Delta_s$ and $\Delta_{\lambda L}$. Through the diagonalization of the $K^{[-2,-1]}$, we get the following expression for the three $U(1)$ charges

\begin{eqnarray}
\Delta_Q&=&\Delta N_{1,\uparrow} + \Delta N_{2,\uparrow} + \Delta N_{1,\downarrow}\label{eq:ChargeQ}\\
\Delta_s&=&\frac{1}{2}\left(\Delta N_{1,\uparrow} + \Delta N_{2,\uparrow}\right) - \Delta N_{1,\downarrow}\label{eq:ChargeS}\\
\Delta_{\lambda L}&=&\Delta N_{1,\uparrow} - \Delta N_{2,\uparrow}\label{eq:ChargeLambda}
\end{eqnarray}

Here $\Delta_Q$ is the usual total electric charge carried by the propagating mode. The two other charges are associated to the two counter-propagating edge mode, $\Delta_{\lambda L}$ is the charge imbalance for spin up between two lambda levels. $\Delta_s$ is related to the variation of the spin projection $\Delta S_z$
\begin{eqnarray}
\Delta S_z&=&\frac{1}{2}\left(\Delta N_{1,\uparrow} + \Delta N_{2,\uparrow} - \Delta N_{1,\downarrow}\right)\label{eq:DeltaSz}\\
&=&\frac{1}{3}\left(4 \Delta_s + \Delta_Q\right)\nonumber
\end{eqnarray}

In each sector, we can easily deduce the lowest energy $E_0$ that can be obtained for the system
\begin{eqnarray}
E_0&=& \frac{1}{6} v_Q \Delta_Q ^2 + \frac{1}{3} v_s \Delta_s ^2 +  \frac{1}{4} v_{\lambda L} \Delta_{\lambda L}^2 \label{eq:EdgeModeLowestEnergy}
\end{eqnarray}
 where $v_Q$, $v_s$ and $v_{\lambda L}$ are the velocities of each mode that we don't need to determine for this discussion. The associated momentum, calculated from Eq.~\eqref{k0}, is given by 
\begin{eqnarray}
k_0&=& \frac{1}{3} \Delta_Q ^2 - \frac{1}{3} \Delta_s ^2 - \frac{1}{4} \Delta_{\lambda L}^2 + \left(2 \Delta N_{1,\uparrow} - \Delta N_{2,\uparrow}\right) \label{eq:EdgeModeLowestEnergyMomentum}
\end{eqnarray}
The last term is a correction due to the reference of momentum for particles in the second Landau level (and is different if we use $N_i$ instead of $\Delta N_i$). Indeed for direct flux, the lowest angular momentum that can be reached in the $m$-th Landau level is $-m$. This can also be understood when writing composite fermions states using conformal field theory. There the operator representing a CF in the second Landau level is a descendant of vertex operator combining both the first and the second Landau level~\cite{Hansson-PhysRevB.76.075347,Hansson-PhysRevLett.98.076801,Hansson-2016arXiv160101697H} (see in particular Eq.~21 in Ref.~\onlinecite{Hansson-PhysRevB.76.075347}). The exact form of the term is chosen to explain the data, as we show below.

With this description in hand, we can explain the lowest energy structure of the real space entanglement spectrum (RSES) since it should mimic the one of the edge mode. We describe in Tab.~\ref{ChargeQuantumNumbers} the first few sectors, the different distributions for $\Delta N_{1,\uparrow}$, $\Delta N_{2,\uparrow}$, $\Delta N_{1,\downarrow}$. The RSES only allow to resolve the charges $\Delta_Q$ and $\Delta S_z$ (and thus $\Delta_s$). Therefore, each RSES shows all the accessible $\Delta_{\lambda L}$ sectors.  We can focus first on $\Delta_Q=0$. We see that there is a single option for the lowest energy at $\Delta S_z=0$ at momentum $K_0=0$. For $\Delta S_z=+1$, we have two options that lead to the same energy since they have the same $\Delta_{\lambda L}^2$. But these two cases have a different momenta respectively $K_0=+1$ and $K_0=-2$ due to the linear term in Eq.~\eqref{eq:EdgeModeLowestEnergyMomentum}. This is exactly what we observe in the RSES. Switching to  $\Delta_Q=1$, we have the same alternation between a single lowest state at $\Delta S_z=-1/2$ and two lowest states at $\Delta S_z=+1/2$. Note that the finite size system at $N=17$, Eq.~\eqref{eq:EdgeModeLowestEnergyMomentum} predicts the correct momenta for the lowest lying entanglement energies for $N_A=8$ with $S_{z,A}=0, 1, 2$  and $N_A=7$ with $S_{z,A}=-1/2, 1/2, 3/2, 5/2$. 

\begin{figure}
\centerline{\includegraphics[width=0.92\linewidth]{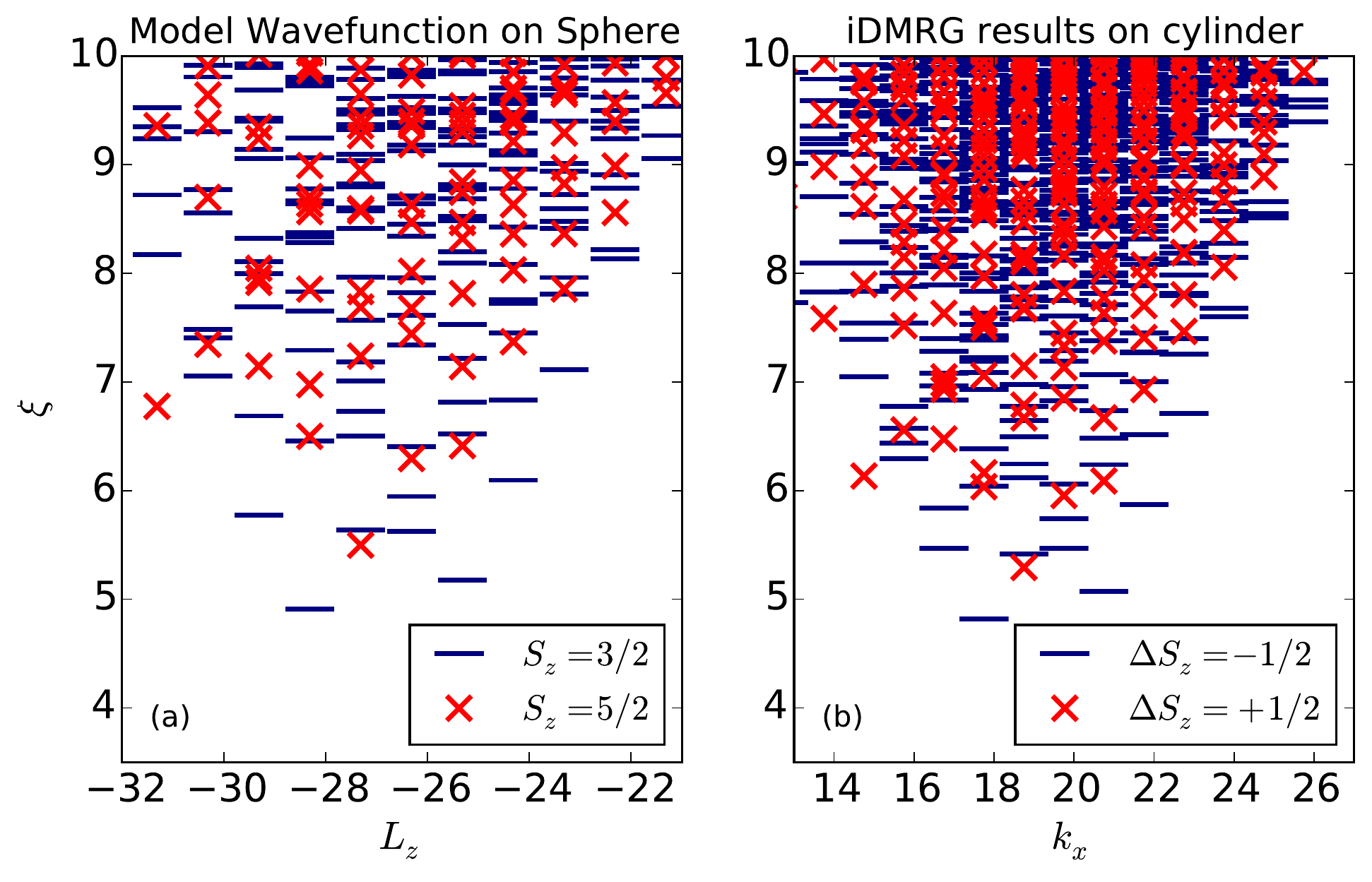}}
\caption{Same as Fig.~\ref{fig:RSES_nu32}, but (a) has $N_A=7$ and (b) has $\Delta N=1$. We again see the expected countings, and the relationship between the various charge sectors is in agreement with the predictions in the lower half of Table \ref{ChargeQuantumNumbers}. }
\label{fig:RSES_nu32_odd}
\end{figure}

A similar analysis can be performed to understand the entanglement spectra at $\nu=4/3$, though in that case since the $K$-matrix is four-dimensional there are two additional charges which need to be summed over, corresponding to the differences in Lambda level occupation for each spin species.

\begin{table}
\begin{tabular}{| c | c | c || c | c | c | c | c |}
\hline
\hline
$\Delta N_{1,\uparrow}$ & $\Delta N_{2,\uparrow}$ & $\Delta N_{1,\downarrow}$ & $\Delta_Q$ & $\Delta_s$ & $\Delta S_z$ & $\Delta_{\lambda L}$ & $k_0$ \\
\hline
\hline
$0$ & $0$  & $0$  & $0$  & $0$  & $0$  & $0$ & $0$ \\
$1$ & $0$  & $-1$  & $0$  & $+3/2$  & $+1$  & $+1$ & $1$\\
$0$ & $1$  & $-1$  & $0$  & $+3/2$  & $+1$  & $-1$ & $-2$\\
\hline
$1$ & $0$  & $0$  & $1$  & $+1/2$  & $+1/2$  & $+1$ & $ 2$ \\
$0$ & $1$  & $0$  & $1$  & $+1/2$  & $+1/2$  & $-1$ & $1$ \\
$0$ & $0$  & $1$  & $1$  & $-1$  & $-1/2$  & $0$ & $0$ \\
\hline
\hline
\end{tabular}
\caption{The different distributions for $\Delta N_{1,\uparrow}$, $\Delta N_{2,\uparrow}$, $\Delta N_{1,\downarrow}$ in the lowest energy sectors.}
\label{ChargeQuantumNumbers}
\end{table}

%%%%%%%%%%%%%%%%%%%%%%%%%%%%%%%%%%%%%%%%%%%%%%%%%%%%%%%%%%%%%%%%%%%%%%%%%%%55
\section{Overlaps in finite size}\label{app:overlaps}

The overlaps with respect to several Jain CF $\Psi^{[-n, -n]}_{\rm CF}$ states were already discussed in great detail in Ref.~\onlinecite{Wu-PhysRevB.87.245123}. In this appendix, we remind the reader of some of the results obtained in that article and provide some additional data by going to slightly higher system sizes. These CF model states are generated by performing the faithful projection onto the lowest Landau level. While rigorous, this approach has the major disadvantage to scale as $N!$ where $N$ is the number of bosons limiting its scope to small systems almost independently of their Hilbert space dimension.  

We start with the CF model state for the bIQHE at $\nu=2$ i.e., $\Psi^{[-1, -1]}_{\rm CF}$. The overlaps with the hardcore interaction ground state are given in Tab.\ref{OverlapCFnu2}. More interestingly, at $\nu=4/3$, we can compute the overlaps for both the NASS state and the $\Psi^{[-2, -2]}_{\rm CF}$. Since we are considering the sphere and since these two states have a different shift, we cannot compute an overlap between them directly. The overlaps with the hardcore interaction ground state for these two model states are shown in Tab.\ref{OverlapNASSCFnu43}. As can be observed, the $\Psi^{[-2, -2]}_{\rm CF}$ has slightly higher overlaps, without completely ruling out the NASS state. Nevertheless, the trend is in agreement with our iDMRG results that clearly favor the CF state. Note that for the NASS state, computing overlaps on the cylinder seems to indicate that the overlap is improved when considering thinner cylinders. This is consistent with the iDMRG that the NASS could emerge for small perimeters and also previous evidence of the NASS state on the torus geometry~\cite{Furukawa-PhysRevA.86.031604}. 
In particular, we used iDMRG to study both the momentum polarization and real space entanglement spectra for cylinders with momenta $L\le10$. Though the small sizes limit the quality of our data, we find a positive shift and a chiral entanglement spectra, both of which are more consistent with a NASS state than the $\wf{-2}{-2}$ state.

\begin{table}
\begin{tabular}{| c | c |}
\hline
$N$ & $\left|\Braket{\Psi_{V_0} | \Psi^{[-1, -1]}_{\rm CF}}\right|^2$\\
\hline
$6$  & $0.9655$\\
$8$  & $0.8197$\\
$10$ & $0.9463$\\
$12$ & $0.8902$\\
$14$ & $0.7886$\\
$16$ & $0.8321$\\
\hline
\end{tabular}
\caption{Overlap between the hardcore interaction ground state for spinful bosons $\Psi_{V_0}$ and the spinful Jain CF state $\Psi^{[-1, -1]}_{\rm CF}$. The largest Hilbert space dimension using only the $S_z$ and $L_z$ quantum numbers and the two discrete symmetries $L_z \leftrightarrow -L_z$ and $S_z \leftrightarrow -S_z$ is $1.2 \times 10^{6}$.}
\label{OverlapCFnu2}
\end{table}

\begin{table}
\begin{tabular}{| c | c | c |}
\hline
$N$ & $\left|\Braket{\Psi_{V_0}|\Psi^{[-2, -2]}_{\rm CF}}\right|^2$ & $\left|\braket{\Psi_{V_0}|\Psi_{\rm NASS}}\right|^2$ \\
\hline
$8$ & $0.9957$ & $0.8457$\\
$12$ & $0.9711$ & $0.8429$ \\
$16$ & $0.9268$ & $0.8054$ \\
\hline
\end{tabular}
\caption{Overlap between the hardcore interaction ground state for spinful bosons $\Psi_{V_0}$ and the spinful Jain CF state $\Psi^{[-2, -2]}_{\rm CF}$ and the NASS state $\Psi_{\rm NASS}$. The overlap is defined as $\left|\Braket{\Psi_{V_0}|\Psi^{[-2, -2]}_{\rm CF}}\right|^2$ on the sphere geometry. Note that the two model wavefunctions for a given number of bosons do not occur at the same number of flux quanta due to a different shift. The largest Hilbert space dimensions using only the $S_z$ and $L_z$ quantum numbers and the two discrete symmetries $L_z \leftrightarrow -L_z$ and $S_z \leftrightarrow -S_z$ are $1.2 \times 10^{7}$ (for the NASS state) and $2.1 \times 10^{8}$ (for the CF state).}
\label{OverlapNASSCFnu43}
\end{table}

\begin{figure}
\centerline{\includegraphics[width=0.62\linewidth]{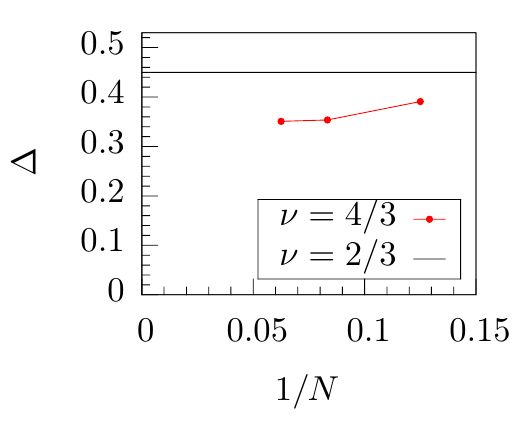}}
\caption{Neutral gaps $\Delta$ for the hardcore interaction at $\nu=4/3$ on the sphere geometry at shift $\delta=-1$. Only three sizes are available $N=8$, $12$ and $16$ (red dots), preventing any finite size extrapolation. The horizontal black line is a guide for the eye showing the thermodynamical extrapolation of the neutral gap above the Halperin $(221)$ state for the hardcore interaction\cite{Sterdyniak-PhysRevB.91.035115}.}
\label{fig:GapNu43}
\end{figure}

\begin{figure}
\centerline{\includegraphics[width=0.92\linewidth]{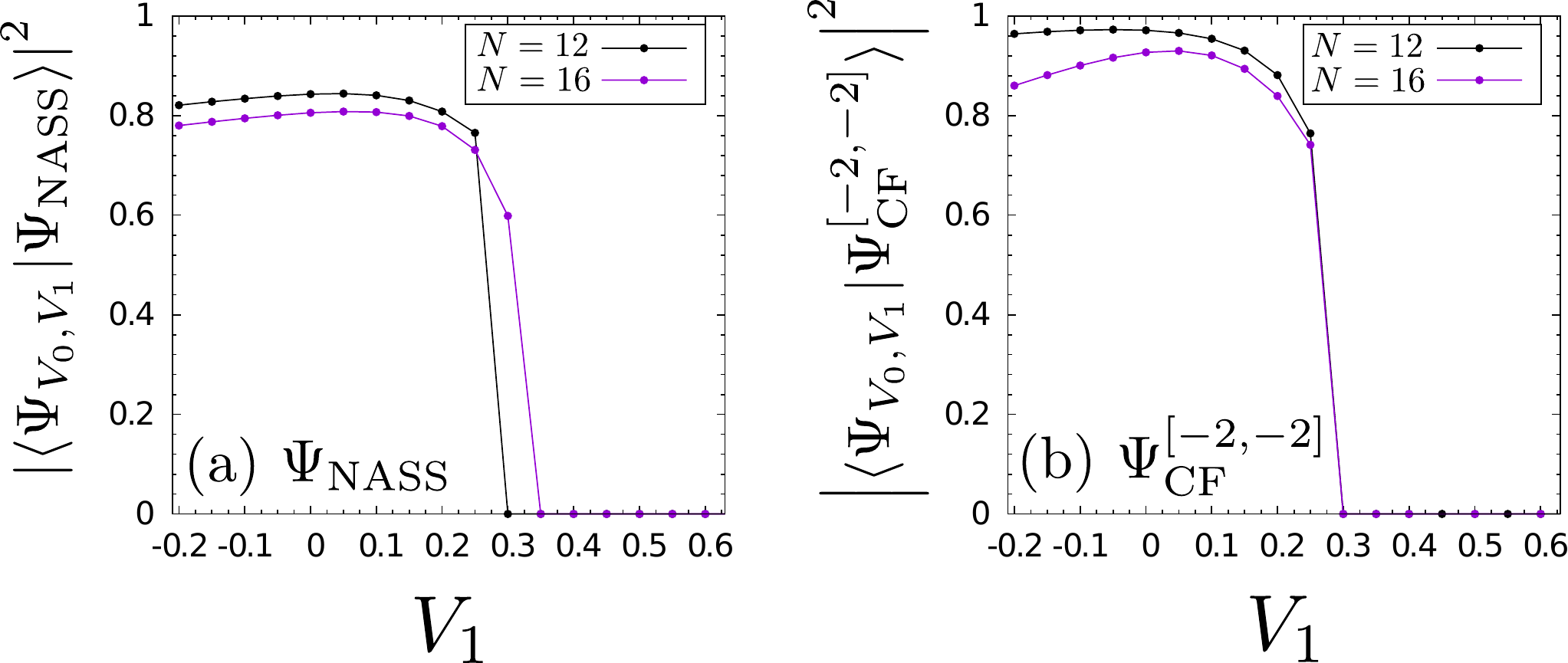}}
\caption{Overlap between the ground state $\Psi_{V_0,V_1}$ of the Hamiltonian using only the two pseudo-potentials $V_0=1$ and $V_1$ and the NASS state (left panel) or the spinful Jain CF state $\Psi^{[-1, -1]}_{\rm CF}$ (right panel). The calculations were performed on the sphere geometry for both $N=12$ (black line) and $N=16$ (purple line) bosons.}
\label{fig:OverlapNu43}
\end{figure}

We can move away from the hardcore interaction by adding some $V_1$ pseudo-potential and see how these overlaps are modified, giving some hint about the stability of these candidate phases. We focus on $\nu=4/3$. The overlaps as a function of $V_1$ are shown in Fig.~\ref{fig:OverlapNu43}. The picture is unchanged, namely the two candidates are comparably stable with respect to $V_1$ with a slight edge for the CF state. In particular, the transition to a fully polarized state occurs around $V_1 \simeq 0.4$ irrespective of the shift. This value of $V_1$ also leads to the system full polarization for $\nu=1$~\cite{Liu-PhysRevB.93.085115}

For spinless fermions, the particle-hole symmetry is valid for the whole spectrum. This implies that the gaps (both charge and neutral) are identical. It is interesting to see if the bosonic PH symmetry, though not microscopic as in the fermionic case, can still relate the gaps of states at, e.g., $\nu = 2/3$ and $4/3$. Here we will focus on the neutral gap. For $\nu=2/3$ and the hardcore interaction, it was numerically evaluated to $\Delta \simeq 0.45$ in Ref.~\onlinecite{Sterdyniak-PhysRevB.91.035115}. For $\nu=4/3$, the situation is more complicated. Due to the competing NASS phase on the torus geometry, we have to focus on the sphere geometry where a suitable choice of the shift can prevent this competition. Moreover, only three system sizes are numerically doable, preventing any thermodynamical extrapolation. The results are shown in Fig.~\ref{fig:GapNu43}. A plausible value for the extrapolated neutral gap would be $0.3 \leq \Delta \leq 0.4$, which is slightly smaller than the one at $\nu=2/3$. Therefore it seems that, at least in this case, the bosonic PH symmetry does not extend beyond the low-energy properties.

\begin{figure}
\centerline{\includegraphics[width=0.45\linewidth]{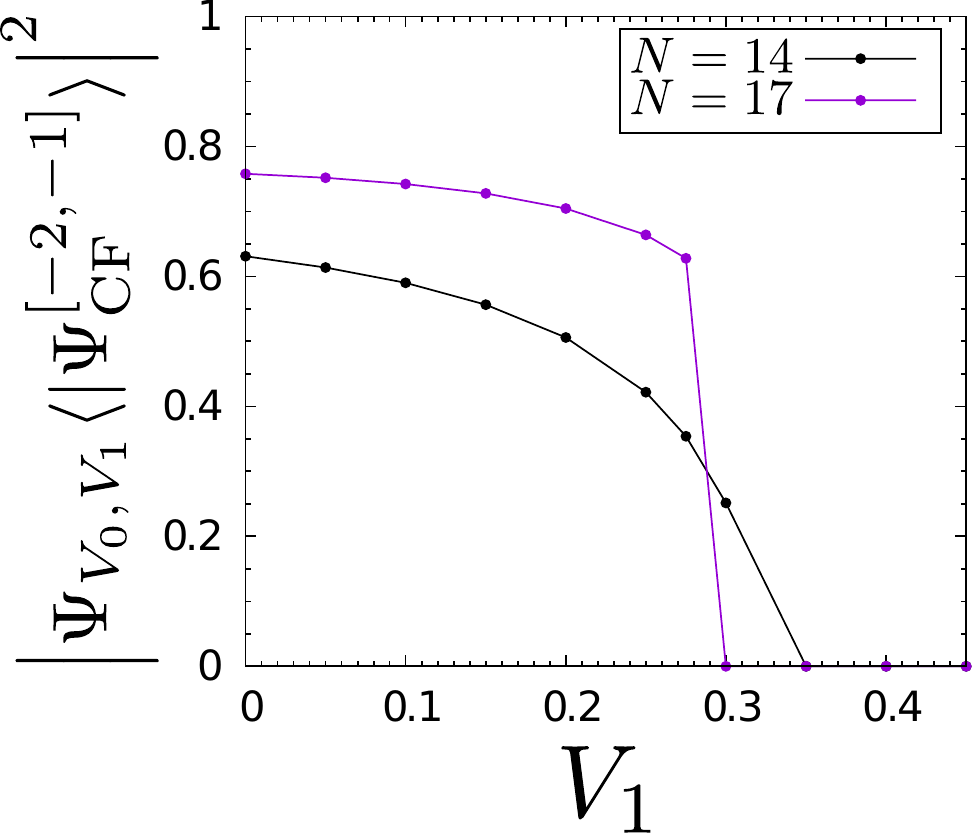}}
\caption{Overlap between the ground state $\Psi_{V_0,V_1}$ of the Hamiltonian using only the two pseudo-potentials $V_0=1$ and $V_1$ and the the spinful Jain CF state $\Psi^{[-2, -1]}_{\rm CF}$ in the spin sector $S=\frac{N}{3}$. The calculations were performed on the sphere geometry for both $N=14$ (black line) and $N=17$ (purple line) bosons.}
\label{fig:OverlapNu32}
\end{figure}

Finally, we address the case of $\nu=3/2$. The Jain CF state $\Psi^{[-2, -1]}_{\rm CF}$ is a partially polarized state with a total spin $S=\frac{N}{3}$ for $N$ bosons. This candidate is not relevant when considering the absolute ground state of the hardcore interaction but has some substantial overlap with the ground state in the total spin sector corresponding to this model state. These overlaps are given in Tab.~\ref{OverlapCFnu32} for the sphere geometry. Note that the lower overlap value for $N=14$ might be due to some aliasing. Tuning the $V_1$ pseudo-potential plays two roles here. It might improve the overlap, and also shows the stability of the model in the polarization sector $S=\frac{N}{3}$. As can be observed in Fig.~\ref{fig:OverlapNu32}, the $V_1=0$ is the optimum case and adding some $V_1$ has a minor effect on the overlap until the system fully polarizes around $V_1 \simeq 0.3$. Second, we might wonder if adding some $V_1$ could drive the system absolute ground state into the wanted total spin sector. What we have observed using exact diagonalizations both on the torus and the sphere geometry is that is mostly occur close to the transition toward a fully polarized system. For example in the cases shown in Fig.~\ref{fig:OverlapNu32}, the absolute ground state has $S=\frac{N}{3}$ between $V_1\simeq 0.2$ and $V_1\simeq 0.3$ for $N=14$. But we never found such polarization at the $V_1$ resolution we have used for $N=17$.

\begin{table}
\begin{tabular}{| c | c |}
\hline
$N$ & $\left|\Braket{\Psi_{V_0}|\Psi^{[-2, -1]}_{\rm CF}}\right|^2$ \\
\hline
$8$ & $0.9130$ \\
$11$ & $0.8900$ \\
$14$ & $0.6309$ \\
$17$ & $0.7576$ \\
\hline
\end{tabular}
\caption{Overlap between the hardcore interaction ground state for spinful bosons $\Psi_{V_0}$ in the spin total spin sector $S$ and the spinful Jain CF state $\Psi^{[-2, -1]}_{\rm CF}$. The overlap is defined as $\left|\Braket{\Psi_{V_0}|\Psi^{[-2, -1]}_{\rm CF}}\right|^2$ on the sphere geometry. The $\Psi^{[-2, -1]}_{\rm CF}$ state has a fractional shift $\delta=-\frac{2}{3}$, leading to particle numbers that are not multiple of $3$. The largest Hilbert space dimensions using only the $S_z$ and $L_z$ quantum numbers and the the discrete symmetry $L_z \leftrightarrow -L_z$ is $1.17 \times 10^{8}$.}
\label{OverlapCFnu32}
\end{table}

%%%%%%%%%%%%%%%%%%%%%%%%%%%%%%%%%%%%%%%%%%%%%%%%%%%%%%%%%%%%%%%%%%%%%%%%%%%55
\section{Additional CFL evidence}\label{app:CFLEvidence}

\begin{figure}[htb]
\includegraphics[width=0.95\linewidth]{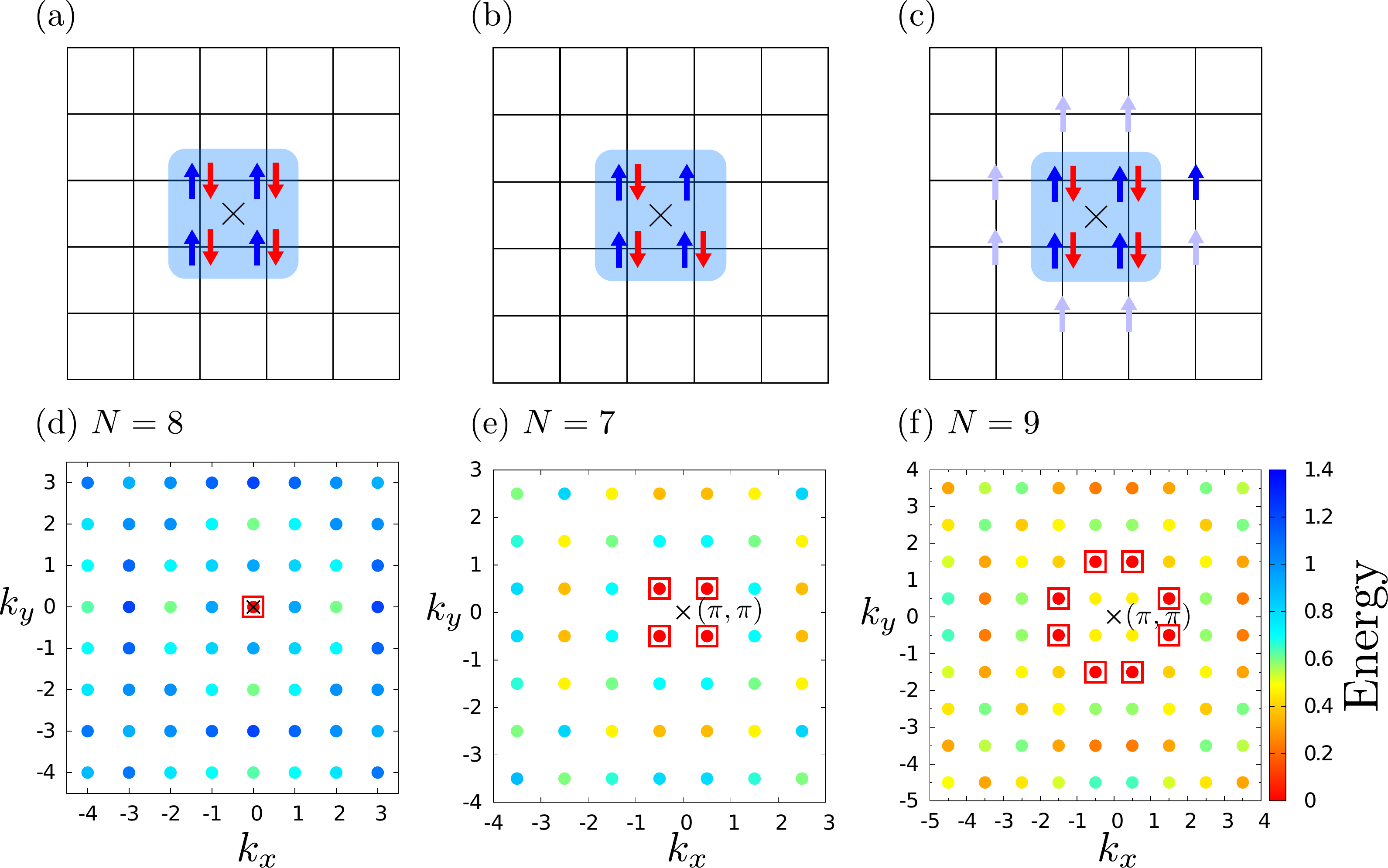}
\caption{{\it Upper panel}: Filling up the Brillouin zone with free spinful composite fermions to form a Fermi sea on a square torus (a, b, c). For $N=8$ (a) there is a unique ground state provided there is a half-flux shift in the position of the origin (denoted by a cross in a, b, c). Removing one composite fermion on the outer shell of the Fermi sea creates a quasihole excitation (b). Adding a composite fermion to one the the sites with the shortest distance to the center of the Fermi sea creates a quasielectron excitation (c). For the quasielectron excitation we depict the equivalent positions in reciprocal space in light blue. {\it Lower panel}: Lowest energies in each momentum sector for the hardcore hamiltonian at $N=N_\Phi$ on a square torus (d, e, f). The lowest energies are indicated by a red box. (d), (e) and (f) involve respectively $N=8$, $7$ and $9$. Plot (d) is centered (black cross) around $(0,0)$ and plots (e) and (f) are centered around $(\pi,\pi)$.}
\label{fig:SquareCFFS}
\end{figure}

\begin{figure}[bt]
\includegraphics[width=0.35\linewidth]{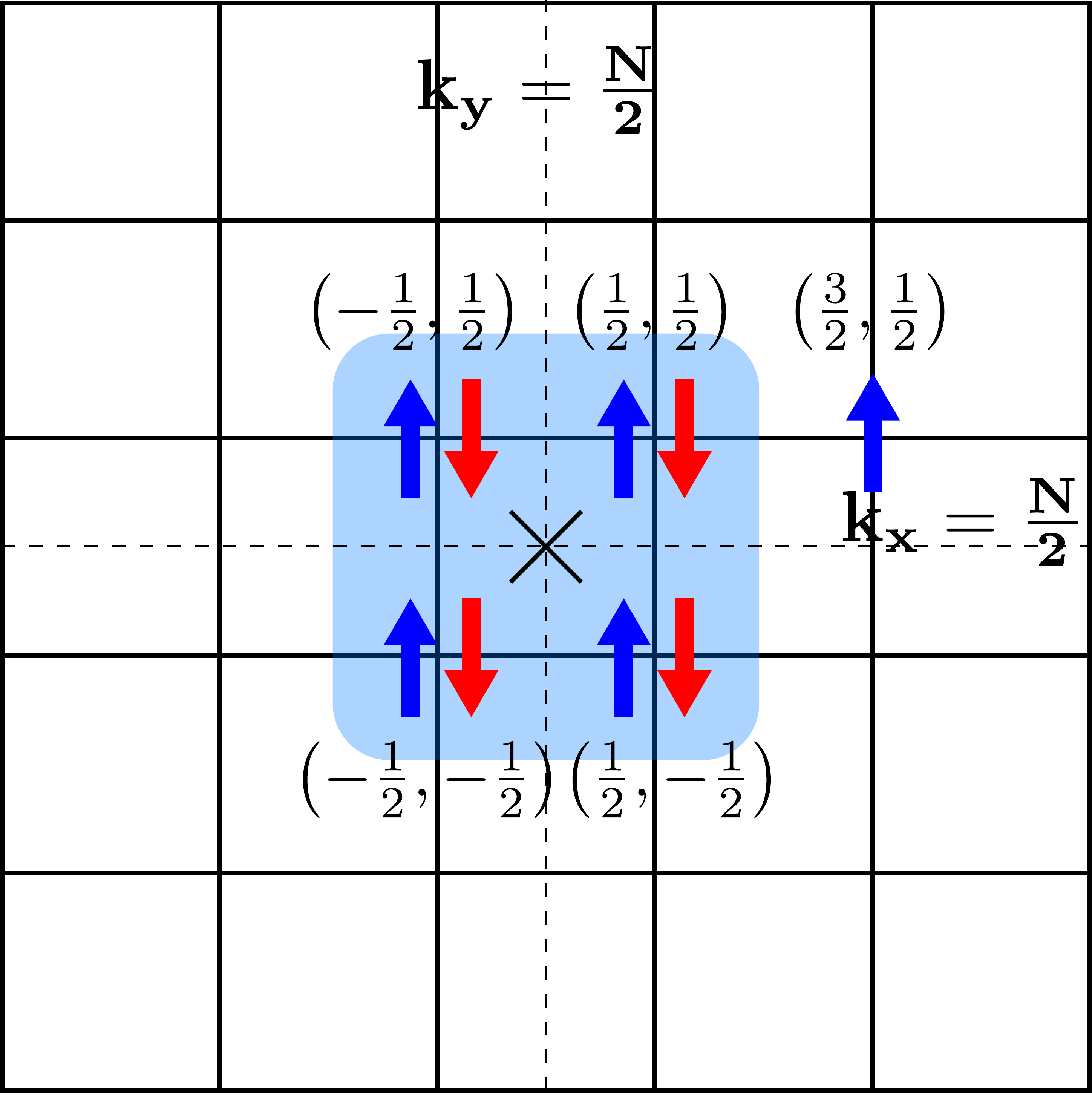}
\caption{Predicting the ground state momentum sector for $N=9$ spinful bosons on a square torus. The origin is depicted by a cross and lies at momentum $\left(\frac{N}{2}, \frac{N}{2}\right)$ in units of $\frac{2\pi}{N}$. The numbers in parenthesis indicate the position of a CF relative to the origin in units of $\frac{2\pi}{N}$.}
\label{fig:squareCFFSmomentum}
\end{figure}

The CFL  construction described in Sec.~\ref{Sec:Nu1} also works for the square torus. On this geometry, we can place the origin of the CF dispersion relation on an accessible point of the Brillouin zone as we did for the $\theta = \pi/3$ torus. But unlike the $C_{6v}$ symmetry, the $C_{4v}$ symmetry also allows for a half-flux shift of the origin in both directions (see Fig.~\ref{fig:SquareCFFS}a). The first option predicts a unique ground state for $N=10$, which is not observed, while the second configuration (Fig.~\ref{fig:SquareCFFS}a) predicts a unique ground state for $N=8$, which we observe in our exact diagonalization data in Fig.~\ref{fig:SquareCFFS}d. Removing (respectively adding) one boson and one flux quantum -- i.e., one CF -- yields a ground state with a degeneracy $4$ (respectively $8$) (see Figs.~\ref{fig:SquareCFFS}b, c). These states appear in our exact diagonalization data as exactly degenerate ground states (they are related by the $C_{4v}$ symmetry) centered around the $(\pi, \pi)$ point in the $N=7$ and $N=9$ spectra as shown in Figs.~\ref{fig:SquareCFFS}e and f. In Fig.~\ref{fig:squareCFFSmomentum}, we explain how the momentum sector of the $N=9$ ground state is predicted. 

\begin{figure}[bth]
\includegraphics[width=\linewidth]{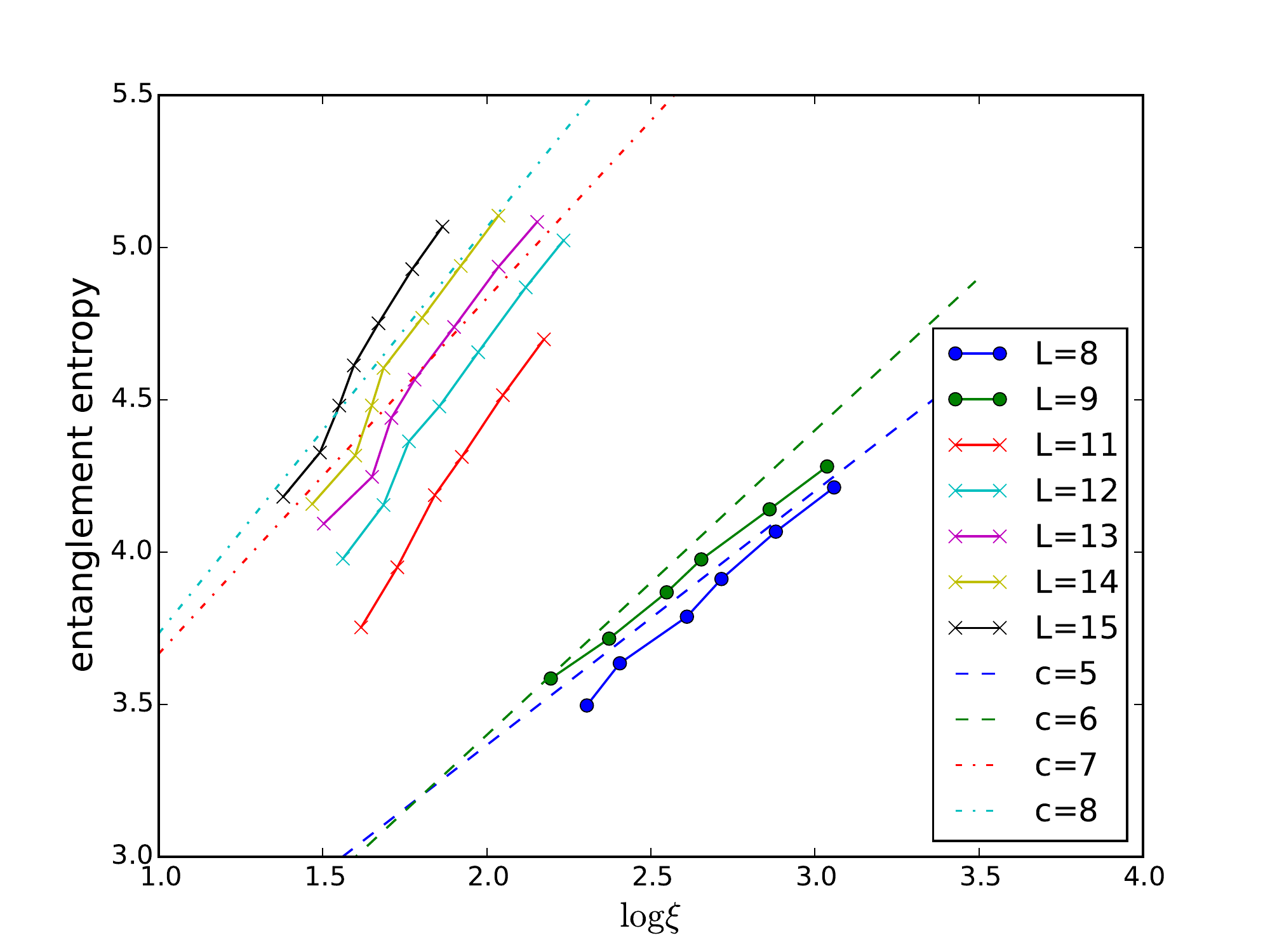}
\caption{Extracting central charge from the slope of entanglement entropy vs. $\log\xi$ (for correlation length $\xi$). The dashed lines show the predictions of $c=5$ and $c=6$ for $L=8-9$ and $c=7$ and $c=8$ for $L=11-15$. Data was taken for bond dimensions in the range $800-5400$.}
\label{central_charge}
\end{figure}

We have also tried to extract the central charge from the iDMRG by plotting entanglement entropy against $\log\xi$, $\xi$ being the correlation length. Such data should be linear with slope $c/6$. As discussed in Ref.~\onlinecite{Geraedts197} the central charge should be given by the total number of wires $-1$. Based on our conclusions about boundary conditions in the main text, we would therefore expect a central charge of $5$ for $L=8-9$, and $7$ for $L=11-15$. We compare these predictions with iDMRG data in Fig.~\ref{central_charge}. The data clearly exhibit a jump of two units for the central charge as soon as the system can accommodate an additional wire. For $L=8-9$ the data matches this prediction (i.e., $c=5$) without completely ruling out a value such as $c=6$. For larger $L$ the slope of the lines seems slightly larger than our predictions. We believe this is because we have not reached large enough bond dimensions. However the clear change in slope, exactly where we have found that the boundary conditions change, is a strong confirmation of our analysis in the main text.

\end{appendix}

\end{document}